\documentclass[journal=jctcce, layout=twocolumns]{achemso}

 \usepackage[latin1]{inputenc}
\usepackage{amsmath}
\usepackage{booktabs}
\usepackage{amstext}
\usepackage{graphicx}
\usepackage{xr}
\usepackage{xcolor}

\usepackage{multirow,colortbl}
\newcommand{\rubpy}{[Ru(bpy)$_{3}$]$^{2+}$}
\newcommand{\rupapy}{[Ru(PaPy$_{3}$)(NO)]$^{2+}$}
\newcommand{\rupy}{[RuPy$_{4}$Cl(NO)]$^{2+}$}
\newcommand{\sharc}{{\sc Sharc}}
\newcommand{\adf}{{\sc Adf}}

\newcommand{\Romannumeral}[1]{\uppercase\expandafter{\romannumeral#1}}
\title{Assessing Excited State Energy Gaps with Time-Dependent Density Functional Theory on Ru(II) Complexes}
\author{Andrew J.~Atkins}
\email{andrew.atkins@univie.ac.at}
\affiliation{Institute of Theoretical Chemistry, Faculty of Chemistry, University of Vienna, W\"{a}hringer Stra{\ss}e 17, A-1090 Vienna, Austria}
\author{Francesco Talotta}
\affiliation{Institute of Theoretical Chemistry, Faculty of Chemistry, University of Vienna, W\"{a}hringer Stra{\ss}e 17, A-1090 Vienna, Austria}
\alsoaffiliation{Laboratoire de Chimie et Physique Quantiques (UMR5626), CNRS et Universit\'{e} de Toulouse 3, Toulouse, France}
\author{Leon Freitag}
\affiliation{Institute of Theoretical Chemistry, Faculty of Chemistry, University of Vienna, W\"{a}hringer Stra{\ss}e 17, A-1090 Vienna, Austria}
\altaffiliation{Present Address: Laboratory of Physical Chemistry, ETH Z\"{u}rich, Vladimir-Prelog-Weg 1-5/10, 8093 Z\"{u}rich,Switzerland}
\author{Martial Boggio-Pasqua}
\affiliation{Laboratoire de Chimie et Physique Quantiques (UMR5626), CNRS et Universit\'{e} de Toulouse 3, Toulouse, France}
\author{Leticia Gonz\'alez}
\email{leticia.gonzalez@univie.ac.at}
\affiliation{Institute of Theoretical Chemistry, Faculty of Chemistry, University of Vienna, W\"{a}hringer Stra{\ss}e 17, A-1090 Vienna, Austria}
\keywords{DFT, \sharc\, Transition Metal, Intersystem Crossing, Photophysics, Photosensitizer}

\AbstractOn
\setkeys{acs}{super = true}
\setcitestyle{super,open={},close={}}

\begin{document}

\begin{tocentry}
\includegraphics[height=3.5cm]{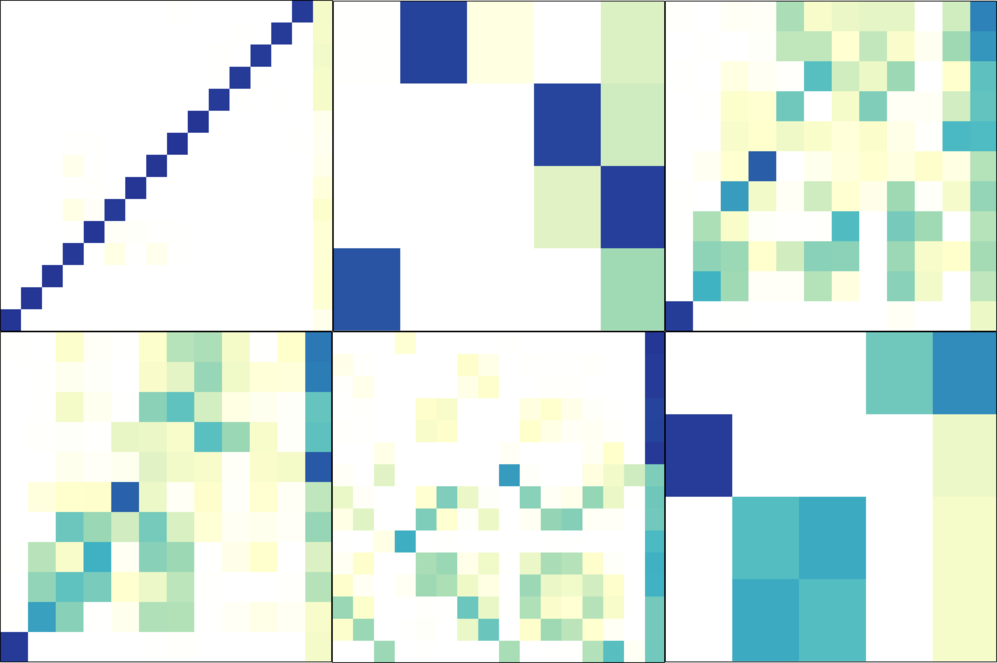}
\end{tocentry}

%
%
%
%
%
%
%
%
%
%
%
%
\newpage
\begin{abstract}
A set of density functionals coming from different rungs on Jacob's ladder are employed to evaluate the electronic excited states of three Ru(II) complexes.
While most studies on the performance of density functionals compare the vertical excitation energies, in this work we focus on the energy gaps between the electronic excited states, of the same and different multiplicity. Excited state energy gaps are important for example to determine radiationless transition probabilities.
Besides energies, a functional should deliver the correct state character and state ordering. Therefore, wavefunction overlaps are introduced to systematically evaluate the effect of different functionals on the character of the excited states.
As a reference, the energies and state characters from multi-state second-order perturbation theory complete active space (MS-CASPT2) are used.
In comparison to MS-CASPT2, it is found that while hybrid functionals provide better vertical excitation energies, pure functionals typically give more accurate excited state energy gaps.
Pure functionals are also found to reproduce the state character and ordering in closer agreement to MS-CASPT2 than the hybrid functionals.
\end{abstract}

\section{Introduction}
Density functional theory (DFT) is considered the modern day workhorse of theoretical chemistry~\cite{cramer_density_2009,orio_density_2009,laurent_td-dft_2013}.
This is because of its low computational cost versus comparitively good accuracy, in comparison to wavefunction methods with a similar cost, such as  Hartree-Fock (HF).
Likewise, its time-dependent variant  (TD-DFT) has seen widespread use for the calculation of the electronic spectra of molecules. \cite{yang_effects_2013,shang_DFT/TDDFT_2014,lestrange_calibration_2015,wang_dft_2014,Kupfer_novel_2012,Nazeeruddin_combined_2005}
However, both, DFT and TD-DFT suffer from the fact that the exact form of the exchange-correlation (XC) functional used in determining the electronic properties is not known and thus, has to be approximated.
As a consequence, there exist many functionals which are parameterized differently for determining specific properties.
As there is not just one class of functionals but many which comprise the different rungs of Jacob's ladder,\cite{perdew_jacobs_2001} approaching a new problem always involves choosing the  correct functional best suited for your system and property of interest.
This requires systematic testing \cite{krewald_redox_2016} and comparison to a benchmark, be it experimental  or coming from very accurate quantum chemical methods.
One limitation of DFT, and of any single-configurational method, is its inability to describe multi-reference systems.
For instance, in many transition metals the issue of multi-reference character is quite prevalent, especially in the 3d elements.
One quintessentially pathological case is Cr$_{2}$ \cite{muller_large-scale_2009}.

For the calculation of excited state properties, the use of TD-DFT is very practical, especially for large molecules,\cite{casida_time_1995} but it
has additional limitations.
A well-know problem is the severe underestimation of charge-transfer excited states \cite{peach_excitation_2008, autschbach_charge-transfer_2009}.
This can be most problematic when using XC functionals based on the Generalized Gradient Approximation (GGA), as the use of HF exact exchange within a functional partially alleviates this error.\cite{dreuw_jcp_2003}
Better improvement is achieved with range-separated functionals like CAM-B3LYP \cite{yanai_new_2004}, which split the two-electron operator into a short- and long-range component.
Rydberg states also tend to be poorly treated by most DFT functionals due to the wrong asymptotic character of the potential far away from the atom by most functionals.
Including HF exact exchange also partially helps as the HF exchange has the correct $\frac{1}{R}$ dependence.
However, the LB94 functional \cite{van_leeuwen_exchange-correlation_1994} is more suitable in this case, since it employs a model potential that fixes the character at long range.
Because the correct asymptotic character is also lost at short range and alters more local properties, the SAOP functional \cite{gritsenko_approximation_1999,schipper_molecular_2000} was designed with a model potential that fixes the asymptotic character of both the long and short range.
The strengths and weaknesses of TD-DFT are discussed in Refs.~\citenum{casida_progress_2012} and \citenum{maitra_perspective:_2016}.
Furthermore, the use of TD-DFT in dynamics is discussed in detail in
Ref.~\citenum{marques_non-bornoppenheimer_2012}.

A number of benchmark studies exist in the literature~\cite{zhao_m06_2008,karton_highly_2008} covering properties for different systems, most often organic molecules where large scale testing over hundreds of molecules is not prohibitively expensive.
One example is the benchmark study of Jacquemin and coworkers on the calculation of excitation energies of a series of small organic molecules with TD-DFT.~\cite{jacquemin_extensive_2009}
Other benchmark studies have focused on the functional performance for indigo\"{i}d dyes \cite{perpete_td-dft_2009}, singlet-triplet vertical energies of organic molecules \cite{jacquemin_assessment_2010}, endoperoxides \cite{martinez-fernandez_can_2011}, etc.
Note also the excellent review of TD-DFT benchmarking performed before 2013 by Laurent et al.~\cite{laurent_td-dft_2013}.
Comparative studies of DFT on transition metal complexes are also abundant in the literature.
A review from Truhlar and Cramer provides an overview of a number of TD-DFT applications to transition metal complexes up to 2009.~\cite{cramer_density_2009}
In Ref.~\citenum{vlahovic_assessment_2015}, a systematic evaluation of the d-d transitions in aquated ion complexes using TD-DFT and Ligand Field DFT is provided. Many of the transition metal complex benchmarks consider a specific collection of transition metal complexes with a given motif.
One example is the analysis of bacteriochlorins by Petit et al.~\cite{petit_absorption_2005}, who also investigated the effects of basis set on TD-DFT for optical excitations in transition metal complexes \cite{petit_predictions_2005}.
Schultz et al.\cite{schultz_benchmarking_2008} studied the effect of functional choice on a series of 3d transition metal atom cations.
Garino et al.\cite{garino_photochemistry_2013} wrote a review on the photochemistry of transition metal complexes calculated using DFT.
Barone and coworkers~\cite{latouche_TD-DFT_2015} investigated the vertical excitation energies of several Pt(II) and Ir(II) complexes with TD-DFT in comparison with the experiment. See also the work of K\"{o}rbel et al.~\cite{korbel_benchmark_2014} on small transition metal complexes comparing TD-DFT with Bethe-Salpeter calculations.
In all of these studies only vertical excitation energies are considered, as these are easiest to compare to experiment and under/overestimations can be fixed with a post energy shift in the spectrum, as, for example, done in X-ray spectroscopy of transition metal complexes.~\cite{atkins_probing_2012, atkins_chemical_2013, atkins_high-resolution_2015, debeer_george_prediction_2008}
In contrast, the assessment of calculated energy gaps between excited states has been mostly overlooked
despite it being significantly affected by the approximation used.
One representative example is the splitting of the pre-edge peaks in acetyl and vinylferrocene K-edge spectra,~\cite{atkins_probing_2012} where the splittings can be overestimated up to 1 eV.
Accurate energy gaps are also important for the correct simulation of non-adiabatic excited state dynamics as the separation between the electronic states drives the probability to populate a lower-lying state.
For example, in trajectory surface-hopping molecular dynamics methods~\cite{barbatti_nonadiabatic_2011}, the energy gap between two electronic states strongly influences the selection of the active state.
Likewise, when dynamical transitions from singlet to triplet states are considered,\cite{richter_sharc:_2011, mai_general_2015}
the size of the singlet-triplet and triplet-triplet gaps is also of paramount importance to promote intersystem crossing.
In the context of radiationless transitions, the nature of the excited states involved within a transition is also relevant.
For example, for intersystem crossing from singlet to triplet states, the character of the wavefunctions will determine the extent of spin-orbit coupling (SOC) between those states in the context of the El-Sayed rules.\cite{lower_triplet_1966,el-sayed_triplet_1968}
A few number  of recent papers have focused on the accuracy of SOCs obtained from different methods.\cite{jovanovic_performance_2017,dinkelbach_assessment_2017, gao_evaluation_2017}

In this paper, we present an extensive TD-DFT benchmark study of vertical singlet, triplet, and singlet-triplet energy gaps of three Ru(II) transition metal complexes, namely, tris-(2,2-bipyridine)ruthenium (II) (\rubpy), (N,N'-bis(2-pyridylmethyl)amine-N-ethyl-2-pyridine-2-carboxamide)nitrosyl ruthenium (II) (\rupapy)
and trans-chloro (tetrapyridine) nitrosyl-ruthenium(II)  (\rupy),  see Scheme \ref{scheme}.
\rubpy\ is an archetype in polypyridyl metal complexes.
Although unsubstituted \rubpy\ is usually not stable enough for long term use in technology, its derivatives are employed as photosensitizers in multiple photonic applications in many different fields ranging from photovoltaics to medicinal therapy.\cite{Kalyanasundaram_CoordChemRev_Photophysics_1982,Kalyanasundaram_CoordChemRev_Applications_1998,Joachim_Nature_Electronics_2000,hagfeldt_dye-sensitized_2010,Gust_FaradayDiscuss_Realizing_2012,Berardi_ChemSocRev_Molecular_2014}
An interesting property of \rubpy\ and its derivatives is its long-lived triplet excited charge-separated state that makes it useful as a photosensitizer, \cite{muller_[rubpy3]2+_2012} and the fact that intersystem crossing occurs in less than 30 fs. \cite{chergui_ultrafast_2015}
\rupapy\ and \rupy\ are two nitrosyl complexes with potential technological\cite{cormary_[rupy_2009,cormary_structural_2012,khadeeva_two-step_2016} and therapeutic applications.\cite{patra_ruthenium_2003}

Here we evaluate the energy gaps between the lowest singlet and triplet excited states and the corresponding state characters of the three Ru(II) complexes using density functionals from different rungs on Jacob's ladder.
Besides providing a data set to assess the performance of different density functionals on excited state properties, the final goal behind this study is to identify the most appropriate functional or set of those with which ultimately one could perform TD-DFT excited state molecular dynamics on these type of transition metal complexes.~\cite{carvalho_nonadiabatic_2015}

\begin{scheme}[h]
    \caption{The three Ru(II) complexes employed in this work A) \rubpy, B) \rupy, C) \rupapy.}\label{scheme}
    \begin{center}
 \includegraphics[width=0.5\textwidth]{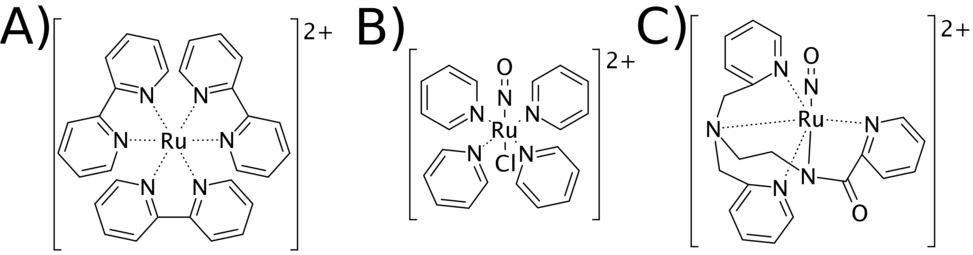}
    \end{center}
\end{scheme}

\section{Computational Methods}
The DFT and TD-DFT calculations on \rubpy have been performed with the \adf\ program package.\cite{guerra_towards_1998, te_velde_chemistry_2001, ADF2014}
The geometry of \rubpy\ was optimised with both the BP86\cite{becke_density-functional_1988, perdew_density-functional_1986} and B3LYP~\cite{becke_density-functional_1993,lee_development_1988,stephens_ab_1994} XC functionals and with and without the D3 correction by
Stefan Grimme \cite{grimme_consistent_2010} using the TZ2P \cite{van_lenthe_optimized_2003} basis set, a small frozen core, scalar relativistics using the zeroth order regular approximation (ZORA)
\cite{van_lenthe_geometry_1999,lenthe_relativistic_1993,van_lenthe_relativistic_1994,van_lenthe_zero-order_1996,van_lenthe_relativistic_1996}, and a Becke \cite{becke_multicenter_1988,franchini_becke_2013} and Zlm fit \cite{franchini_accurate_2014} grid of good quality. In all optimization calculations symmetry was not used.
The D3 correction was tested as there are examples in the literature \cite{jager_using_2015} where it has a very large effect on the optimized structure.
However, in the case of \rubpy\ the use of D3 with or without the Becke-Jones damping \cite{grimme_effect_2011} did not alter the final structure and the structure obtained with both XC functionals were identical.
All symmetry labels are assigned through the use of the completely symmetrized enantiomer ($D_3$ point group).
Nevertheless, the reported values are for the geometry optimized without symmetry constraints.

For \rubpy\ the choice of XC functionals within TD-DFT and their effects on the excited state energies, excited state energy gaps and the state character were analysed on the BP86 optimized geometry.
In these tests a TZ2P basis set was used with a small frozen core, a good quality Becke integration grid and normal quality Zlm spline fit grid.
Scalar relativistics effects were included with ZORA and the SOCs were calculated using a perturbative method, as developed and implemented by Wang and Ziegler \cite{wang_simplified_2005}.
A total of 15 singlet and 15 triplet excitations  using the Davidson iterative procedure was used for the diagonalization.
These states are all of metal-to-ligand charge-transfer (MLCT) states.
Metal-centered (dd) or ligand field states can quench the triplet MLCT luminiscence and therefore are a subject of recurring interest.~\cite{Campagna2007,SUN201587,Hauser2016,C6SC01220E}.
In bare \rubpy, metal-centered states are higher in energy than MLCT states.~\cite{muhavini_wawire_density-functional_2014}

The functionals used to calculate excited states were four GGA XC functionals (BP86, PBE\cite{perdew_generalized_1996}, mPBE\cite{adamo_physically_2002,perdew_generalized_1996}, and BLYP\cite{becke_density-functional_1988,lee_development_1988,johnson_performance_1993,russo_density_1994}), three hybrid XC functionals (PBE0\cite{ernzerhof_assessment_1999,adamo_toward_1999}, BHandHLYP, and B3LYP), two metaGGA XC functionals (TPSS\cite{tao_climbing_2003,staroverov_comparative_2003} and M06-L\cite{zhao_new_2006,zhao_m06_2008}),
two metaHybrid XC functionals (TPSSH\cite{tao_climbing_2003,staroverov_comparative_2003} and M06\cite{zhao_new_2006,zhao_m06_2008}), two Model XC functionals (LB94 and SAOP) and a range-separated functional (CAMY-B3LYP\cite{seth_range-separated_2012, yanai_new_2004,ekstrom_arbitrary-order_2010}).
Moreover, the effect of using the Tamm-Damcoff approximation (TDA) \cite{hirata_time-dependent_1999} was considered for both hybrid and GGA XC functionals.

Along with the XC functional test, other parameters the calculation is dependent on were tested on \rubpy.
These included four different basis sets (QZ4P, TZ2P, TZP and TZP(Ru)-DZP) and three solvation options (COSMO1, COSMO2, COSMO3).
COSMO1 uses the conductor like screening model (COSMO) \cite{pye_implementation_1999,klamt_cosmo:_1993,klamt_conductor-like_1995,klamt_treatment_1996} with an infinite dielectric and a solvent radius of 1.93, COSMO2 uses the values for water (dielectric constant of 78.39 and a solvent radius of 1.93), and COSMO3 uses the values for acetonitrile (dielectric constant of 37.5 and a solvent radius of 2.76).
For all COSMO calculations the Allinger radii \cite{allinger_molecular_1994} were used for the construction of the surface (solvent-excluding-surface \cite{pascual-ahuir_gepol:_1994}).
For the response of the COSMO model we use equilibrium linear-response where the dielectric constant is not frequency dependent, as is the default in ADF.
This entails that the solvent has had time to reorganise and thus relax in the excited states.
Note that one could also employ non-equilibrium linear-response, which means that the solvent is not fully in equilibrium with the excited state density.
This involves a dielectric constant at optical frequencies of 1.77 for water instead of 78.39.
This small dielectric constant represents only the fast response of the electrons of the water.
Thus, all other degrees of freedom of the water are in equilibrium with the ground state density of the solute.
For the other details of COSMO the defaults of \adf\ were used.

As additional parameters, the effect of the choice of integration and fit grid quality was checked, using normal, good, very good and excellent for both the Zlm spline fit and Becke grid.
An alternative, but older, fitting method to the Zlm spline fit is the STO fit method in \adf\ which is tested as well. Also one may opt to use no fitting and thus, utilise the exact density in the calculation. The effect of using the exact density compared to the fitted density using the Zlm spline fit was also tested. Finally, the three options (none, small, and large) for the frozen-core approximation used in \adf~were tested.

For the \rupy\ and \rupapy\ the TD-DFT calculations were done utilising the Gaussian09 program package \cite{g09RevD.01} because it allows us to assess the performance of other XC functionals than those employed in \rubpy.
The geometry of \rupy\ and \rupapy\ at the B3LYP including Grimme's dispersion correction and BP86 levels of theory, were taken from previous work,  Refs.~\citenum{garcia} and \citenum{freitag_theoretical_2014}, respectively.
The subsequent TD-DFT single point calculations\cite{TD_1,furche_adiabatic_2002,TD_3} were performed on both molecules using a cc-pVDZ \cite{cc_1,cc_2} basis set on the hydrogens, a cc-pVTZ \cite{cc_3} basis set for carbon, oxygen and nitrogen.
For ruthenium the Stuttgart relativistic core potential including 28 electrons was used with its associated basis set \cite{core-pot} and two f and one g polarization functions \cite{pol-funcs}.
An ultrafine integration grid was used and no symmetry was employed.
All TD-DFT calculations were performed in both complexes once with and once without TDA.~\cite{hirata_time-dependent_1999,TDA_2}
For \rupy\ 4 singlet and 4 triplet excited states were calculated and for \rupapy\ 10 singlet and 10 triplet excited states were calculated.
The functionals used to calculate the excited states were one LDA (HFS \cite{HFS_1,HFS_2,HFS_3}), four GGA XC functionals (PBE\cite{perdew_generalized_1996,PBE_2}, BP86\cite{perdew_density-functional_1986}, BLYP\cite{becke_density-functional_1988,lee_development_1988,russo_density_1994}, N12\cite{N12}), four metaGGA functionals (TPSS\cite{tao_climbing_2003}, M06-L, MN12-L\cite{MN12L}, SOGGA11\cite{SOGGA11}),
nine Hybrid functionals (B3LYP, BHHLYP, PBE0 \cite{ernzerhof_assessment_1999,adamo_toward_1999}, B971\cite{B971}, B972\cite{B972}, BHH\cite{BHandH-BHandHLYP,lee_development_1988}, HISSbPBE\cite{HISSbPBE}, HSEH1PBE\cite{HSEH1PBE_1,HSEH1PBE_2,HSEH1PBE_3,HSEH1PBE_4,HSEH1PBE_5,HSEH1PBE_6,HSEH1PBE_7}, N12SX\cite{MN12SX-N12SX}), four range-separated
functionals (CAM-B3LYP\cite{yanai_new_2004}, wB97XD\cite{wB97XD}, LC-wPBE\cite{LC-wPBE_1,LC-wPBE_2,LC-wPBE_3}, M11\cite{M11}), and seven metaHybrid functionals (TPSSH\cite{staroverov_comparative_2003,TPSSh_2_Erratum}, M06\cite{zhao_m06_2008}, M06-2X\cite{zhao_m06_2008}, M06HF\cite{M06HF_1,M06HF_2}, MN12SX\cite{MN12SX-N12SX},
tHCTHhyb\cite{tHCTHhyp}, BMK\cite{BMK}).

In order to assess the performance of the XC functionals, a comparison with multi-state complete active space second order perturbation theory (MS-CASPT2)\cite{andersson_second-order_1990,andersson_second-order_1992,finley_multi-state_1998,
Aquilante_JChemTheoryComput_Cholesky_2008} was done for the three complexes.
Although MS-CASPT2 has weaknesses and  one should exercise caution when using it as a reference --as it depends strongly on the choice of the active space or on the so-called IPEA shift~\cite{ghigo_modified_2004,zobel_IPEA_2017}-- for transition metal complexes of this size other accurate options, such as the algebraic diagrammatic construction method to third-order ADC(3) are not computationally feasible.~\cite{plasser_high-level_2015}.
On the positive side, MS-CASPT2 can deal with static and dynamic correlation for arbitrary states and regardless the spin, and is therefore expected to be more consistent than TD-DFT.
For organic molecules, deviations between TD-DFT and MS-CASPT2 and experimental values can be as small as 0.3 eV~\cite{schreiber_benchmarks_2008,silva-junior_benchmarks_2008,zobel_IPEA_2017} or slightly larger\cite{hoyer_multiconfiguration_2016}.
It is also noticeable that TD-DFT tends to provide less accurate values for triplets than for singlet states.
Hence, since we need a one-to-one energy comparison and experimentally it is not possible to have a complete list of energies for the relevant low-lying singlet and triplet states, we resort to employ MS-CASPT2 as a theoretical reference.

\begin{figure}[h]
    \caption{Active orbitals included in the MS-CASPT2 calculation of \rubpy\ (A) and \rupy\ (B)}\label{fig:CASPT2orb}
    \begin{center}
\includegraphics[width=\textwidth]{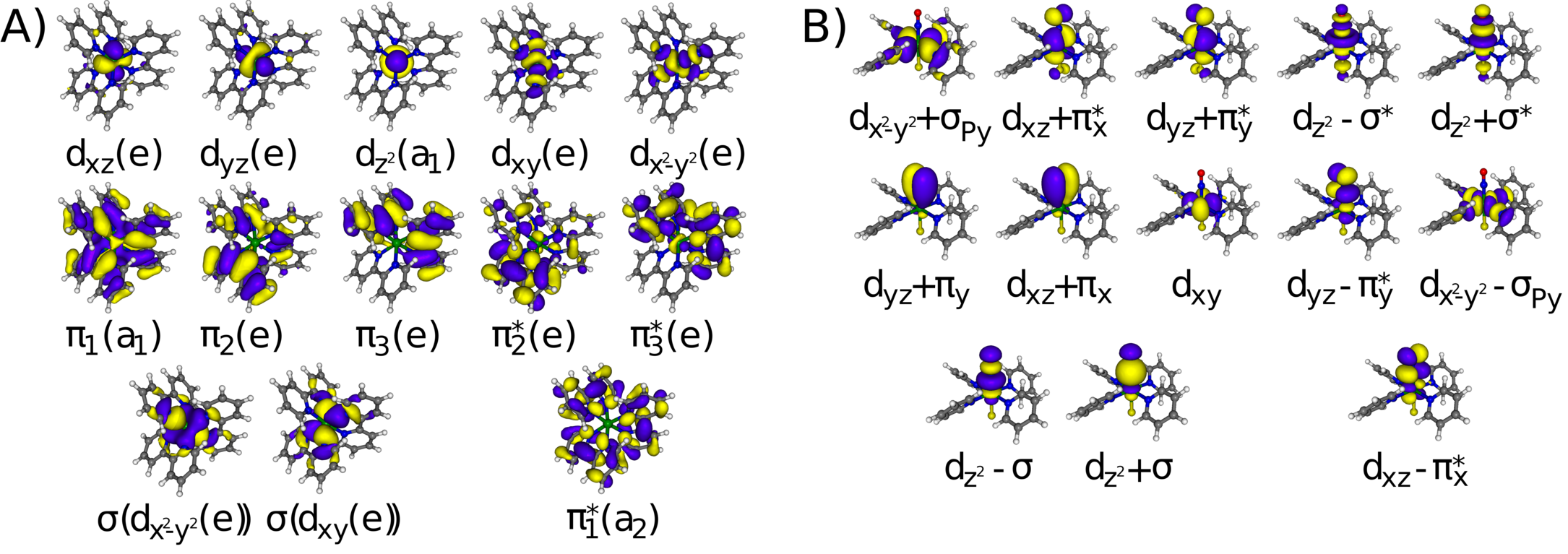}
    \end{center}
    \end{figure}

In the MS-CASPT2 calculations, \rubpy\ was optimized using the wavefunction method, RI-MP2\cite{weigend_ri-mp2:_1997,weigend_ri-mp2:_1998} combined with the def2-TZVP basis set\cite{weigend_balanced_2005} for all atoms, and using $D_3$ symmetry, within the TURBOMOLE 6.5\cite{TURBOMOLE6.5} program package.
The active space of the underlying complete active space self-consistent field (CASSCF) calculation comprised 16 electrons in 13 orbitals (cf.\ Fig.~\ref{fig:CASPT2orb}A): five Ru $4d$ orbitals, two bonding linear combinations of $n$ orbitals with $d_{x^2-y^2}$ and $d_{xy}$ orbitals (labeled $\sigma$($d_{x^2-y^2}$) and $\sigma$($d_{xy}$)) and three $\pi,\pi^*$ orbital pairs -- two of $e$ symmetry and one of $a_1$ and $a_2$ symmetry.
Unlike $3d$ transition metals\cite{pierloot_caspt2_2003,radon_electronic_2010}, Ru does not require the double-shell $d$ orbitals in the active space, if the $\sigma$($d_{x^2-y^2}$) and $\sigma$($d_{xy}$) orbitals are present\cite{escudero_raspt2/rasscf_2012,Freitag_PhysChemChemPhys_Orbital_2015}.
Extensive testing indicated that this is the smallest active space that can be used in the MS-CASPT2 calculations for the chosen number of states. Smaller active spaces, such as (12,9), led to unsatisfactory results.

The ANO-RCC all-electron basis set\cite{roos_new_2005} was used with triple-zeta (ANO-RCC-VTZP) contraction on the Ru atom and double-zeta (ANO-RCC-VDZP) contraction on other atoms.
The use of a smaller basis set, i.e. ANO-RCC-VDZP on Ru and all H atoms and ANO-RCC-VDZP  without d functions on C and N resulted only in an average deviation of 0.14 eV with respect to the larger one. Since this value is smaller than other CASPT2 errors found in the literature, see e.g. Refs.~\citenum{schreiber_benchmarks_2008} and~\citenum{hoyer_multiconfiguration_2016}, we consider this basis set a reasonable choice.
\rubpy\ has D$_3$ symmetry and a straightforward CASSCF wavefunction optimization in C$_1$ symmetry led to a broken-symmetry solution. To ensure the correct symmetry of the wavefunction, orbital rotations between different irreducible representations of the D$_3$ point group were disabled in the CASSCF orbital optimization.

In the MS-CASPT2 calculation, an imaginary level shift~\cite{forsberg_multiconfiguration_1997} of 0.3 a.u.~was used to avoid intruder states.
The standard IPEA shift~\cite{ghigo_modified_2004} of 0.25 a.u.~was used as a dogma, although it is not clear whether this value of the IPEA shift is optimal or not for the present case.
For organic molecules it has been shown that the IPEA shift is certainly not justified. \cite{zobel_IPEA_2017}
On the contrary, in some transition metal complexes an IPEA value of 0.25 a.u.~or larger do improve the agreement with experiment.\cite{kepenekian_energetics_2009,suaud_light-induced_2009,lawson_daku_accurate_2012,vela_zeroth-order_2016,rudavskyi_computational_2014,pierloot_relative_2008,pierloot_relative_2006,kepenekian_what_2009}

Eleven roots were calculated for both the singlet and triplet multiplicities with equal weights.
Two-electron integrals were approximated with the atomic compact Cholesky decomposition (acCD) approach\cite{Aquilante_JChemPhys_Atomic_2009,aquilante_accurate_2008}
with a decomposition threshold of 10$^{-4}$ a.u.
The calculation has been performed with the MOLCAS 8.0 program\cite{Aquilante_JComputChem_Molcas_2015}.

For \rupy\ the ground state geometry was retrieved from Ref.~\citenum{garcia} .
The MS-CASPT2 single-point energies have been obtained with the MOLCAS 8.0 program package\cite{Aquilante_JComputChem_Molcas_2015} using the ANO-RCC-VTZP basis set\cite{roos_new_2005} for the ruthenium atom and the ANO-RCC-VDZP for all other atoms. In all calculations the Cholesky decomposition\cite{Aquilante_JChemPhys_Atomic_2009,aquilante_accurate_2008} was used with a threshold of 10$^{-4}$ a.u.
Seven roots were calculated for both the singlet and triplet states with equal weights.
The imaginary level shift was 0.3 a.u.~and the IPEA shift 0.25 a.u.
The active space, shown in Figure \ref{fig:CASPT2orb}B consists of 16 electrons in 13 orbitals, including the five Ru 4d valence orbitals, the pairs of $\pi$ and $\pi^*$ on the NO ligand together with $\sigma$ and $\sigma^*$ orbitals, and one pair $d\sigma$ and $d\sigma^*$ orbitals of the equatorial pyridine ligands ($\sigma_{py}$).

The MS-CASPT2 results of \rupapy\ are taken from Ref.~\citenum{freitag_theoretical_2014}.
The active space contained 18 electrons in 14 orbitals, including the five Ru 4d orbitals, two pairs of $\pi$
and $\pi^*$ orbitals in the NO ligand, two $\sigma$ orbitals forming the bonding--antibonding pairs with the Ru $d_{x^2-y^2}$ and $d_{z^2}$ orbitals, respectively, and one $\pi$ and $\pi^*$ pair located at the amide moiety.\cite{freitag_theoretical_2014}
Five singlets and six triplet states were averaged in the MS-CASPT2 procedure.
The imaginary level shift was 0.3 a.u.~and the IPEA shift 0.25 a.u.

We note here that in all three cases, the number of roots is not chosen to simulate the full absorption spectrum, but rather to describe the lowest-lying excitations, where interesting photochemical events take place.
In \rubpy, intersystem crossing occurs after excitation at 400 nm (3.10 eV).\cite{chergui_ultrafast_2015}
In \rupy, NO isomerism occurs after irradiation at 473 nm (2.62 eV), which corresponds to the S$_{1}$ and S$_{2}$.\cite{garcia}
Finally in \rupapy, the inclusion of 10 states covers the excited states that lead to dissociation.\cite{freitag_theoretical_2014}

\section{Results}
\subsection{\rubpy}
The following discussion includes the effect of the different density functionals on vertical excitation energy and energy gaps, as well state characters in comparison to the results obtained from the MS-CASPT2 benchmark calculation.
Additionally, a short analysis of the effects of the basis set, solvation models and technical settings is included.

\subsubsection{Density Functional Dependence}
\label{sec:DFTdep}

To evaluate the validity of the MS-CASPT2 benchmark, we use experimental values, even if we do not consider vibronic effects and spin-orbit couplings for simplicity.
Experimental information on \rubpy\ in vacuum is limited\cite{kirketerp_absorption_2010, Xu_electronic_2016} and most measurements have been performed in solution.\cite{chergui_ultrafast_2015}
In gas phase, the only available experimental spectrum~\cite{kirketerp_absorption_2010} shows a very broad band measured at 2.88 eV, while in CH$_{3}$CN reports a maximum of the same band at 2.74 eV and a second shoulder can be resolved at approximately 2.94 eV through
extracting the experimental spectra.
In aqueous solution the MLCT band has been reported to have excitations at 2.74 eV and 2.87 eV \cite{Xu_electronic_2016} or 2.73 eV and 2.90 eV. \cite{heully_spin-orbit_2009}
Both, in gas and solution, these states have been assigned as MLCT states.

As explained before, in order to provide a one-to-one state comparison against TD-DFT, a theoretical calculation with MS-CASPT2(16,13) has been performed.
An additional advantage of comparing calculated values is that possible error cancellations due to experimental conditions not included in the calculation are excluded.
Table \ref{tab:bipy-exc} collects the spin-free excitations energies, state characters and oscillator strengths of the first 8 singlets and 9 triplets of  \rubpy\ calculated with MS-CASPT2(16,13).

\def\mr#1{\multirow{-2}{*}[1pt]{$\left.\rule{0cm}{2.5ex}\right\}$#1}}
\begin{table}[h]
\caption{Symmetries, characters, spin-free MS-CASPT2(16,13), BP86 and B3LYP excitation energies
(in eV) and oscillator strengths $f$ of the lowest singlet and triplet
excited states of \rubpy.}
\label{tab:bipy-exc}
\centering
\begin{tabular}{cccccccccc}
\toprule
&\multicolumn{3}{c}{MS-CASPT2(16,13)}&\multicolumn{3}{c}{BP86}&\multicolumn{3}{c}{B3LYP}\\
     State      & Sym.      & $\Delta E$ & $f$ & Sym.      & $\Delta E$ & $f$ & Sym.      & $\Delta E$ & $f$\\
     \midrule
     $S_1$ & 1$^1A_2$  & 2.58  &0.006 & 1$^1A_2$  & 1.83  &0.001& 1$^1A_2$  & 2.47  &0.001\\\\
     $S_2$ &           & 2.75 &0.010 &           & 2.00 &0.000&&2.49 &0.000\\
     $S_3$ &\mr{1$^1E$}& 2.76 &0.002&\mr{1$^1E$}& 2.01 &0.000&\mr{1$^1E$}& 2.49 &0.000\\
    $S_4$ &           & 2.83     &0.205&           & 2.18     &0.018&1$^1A_2$           & 2.66     &0.000\\
    $S_5$ &\mr{2$^1E$}& 2.86     &0.213&\mr{2$^1E$}& 2.19     &0.018 &      & 2.68     &0.013\\
    $S_6$ & 1$^1A_2$  & 3.04 &0.088& 1$^1A_2$  & 2.27 &0.000 &\mr{2$^1E$}   & 2.69 &0.013\\
    $S_7$ &           & 3.13     &0.200&           & 2.46     &0.057&           & 2.86     &0.106\\
    $S_8$ &\mr{3$^1E$}& 3.14     &0.171&\mr{3$^1E$}& 2.46     &0.057&\mr{3$^1E$}& 2.86     &0.107\\
     \cmidrule(rl){1-10}
      $T_1$ & 1$^3A_2$  & 2.64 $^a$ &--& 1$^3A_2$  & 1.79 &--&   &2.33 &-\\
      $T_2$ &           & 2.71 &-- &           & 1.90 &--&\mr{1$^3E$}           & 2.33 &--\\
      $T_3$ &\mr{1$^3E$}& 2.72 &--&\mr{1$^3E$}& 1.90 &--&1$^3A_2$& 2.36 &--\\
      $T_4$ &           & 2.82 &--&           & 2.09 &--& 1$^3A_1$           & 2.38 &--\\
      $T_5$ &\mr{2$^3E$}& 2.85 &--&\mr{2$^3E$}& 2.09 &-- && 2.48 &--\\
      $T_6$ & 1$^3A_1$  & 2.92 &--& 1$^3A_1$  & 2.10 &--&\mr{2$^3E$}  & 2.48 &--\\
      $T_7$ & 2$^3A_2$  & 3.05 &--&  & 2.18 &--&  & 2.59 &--\\
      $T_8$ &           & 3.13 &--&\mr{3$^3E$}           & 2.18 &--&\mr{3$^3E$}           & 2.59 &--\\
      $T_9$ &\mr{3$^3E$}& 3.18 &--&2$^3A_2$ & 2.24 &--&2$^3A_2$ & 2.62 &--\\
      \bottomrule
\end{tabular}

$^a$ Note that the T$_1$ state is incorrectly 0.05 eV above the S$_1$ state, due to an artificial mixing between T$_1$ and T$_3$, see text.
\end{table}

If one would convolute all the 8 singlet excitations with Gaussians with a full width half maximum (FWHM) of 0.44 eV, a band with a maximum at 2.98 eV would be obtained.
Despite without spin-orbit and vibronic effects a one-to-one comparison is not fair, within these limitations only a deviation of 0.1 eV with respect to the gas phase experimental result of 2.88 eV~\cite{kirketerp_absorption_2010} is obtained.

If we analyse in detail the gas phase spectrum, the intensity of the broad band comes from two overlapping bands rooted at the two 2$^1E$ and 3$^1E$ bright electronic states, which in the employed C$_1$ symmetry, correspond to the pairs (S$_4$,S$_5$) and (S$_7$,S$_8$).
The energy difference between both states, predicted at 2.84 eV and 3.14 eV, respectively (average of both excitations), is 0.3 eV --i.e. \textit{ca.} 0.1 eV larger than in the solvated experiment, where the splitting is 0.17$\pm0.4$ eV (average of three reported differences in solvation
with maximum deviation as error). \cite{heully_spin-orbit_2009,kirketerp_absorption_2010,Xu_electronic_2016}
Despite the small differences in energy that can be attributed to solvent interactions, we consider that the agreement between experiment and theory
is reasonable within the approximations made, validating the use of MS-CASPT2 as a reference for TD-DFT.
Accordingly, the MS-CASPT2 values will be used to analyse the performance of different XC functionals in describing electronic excited states in \rubpy.
First we consider the excitation energies of \rubpy\ with respect to the electronic ground state.

We note that the MS-CASPT2(16,13) calculation on \rubpy gives an artificial mixing of the T$_{1}$ and T$_{3}$ states due to symmetry breaking: despite the D$_3$ symmetry constraint on the wavefunction optimization the A$_2$ T$_1$ and the E T$_3$ states show symmetry-forbidden mixing.
As a consequence, the energies and characters of these states are not fully reliable; for instance, the energy of the T$_1$ state is 0.05 eV above that of the S$_1$ (see Table 1).
Unfortunately, this issue could not be fixed by using larger active spaces, tighter convergence criteria or different basis sets.
However, the symmetry breaking only occurs in these two states, whilst the other states included here are all unaffected.
Hence, we shall exclude the T$_1$ and T$_3$ from our analysis in the following.

The calculated vertical excitation energies with BP86 and B3LYP are also shown in Table \ref{tab:bipy-exc}.
BP86 severely underestimates the excitation energies; for the S$_{1}$ this is underestimated by 0.75 eV and the T$_{2}$ by 0.81 eV.
The inclusion of HF exact exchange in B3LYP seems to reduce the underestimation of the excitation energies compared to MS-CASPT2, with the S$_{1}$ being underestimated by 0.11 eV and the T$_{2}$ by 0.38 eV.
In both cases the triplets are further underestimated than the singlet excited states.

\begin{figure}
\caption{Mean signed errors (MSE) and mean absolute errors (MAE) of the excitation energies (panel A) and the energy gaps for singlets (S) and triplet (T) states (panel B) of \rubpy\ computed against MS-CASPT2(16,13). }
 \label{fig:excenrgfunct}
\begin{center}
 \includegraphics[width=0.6\textwidth]{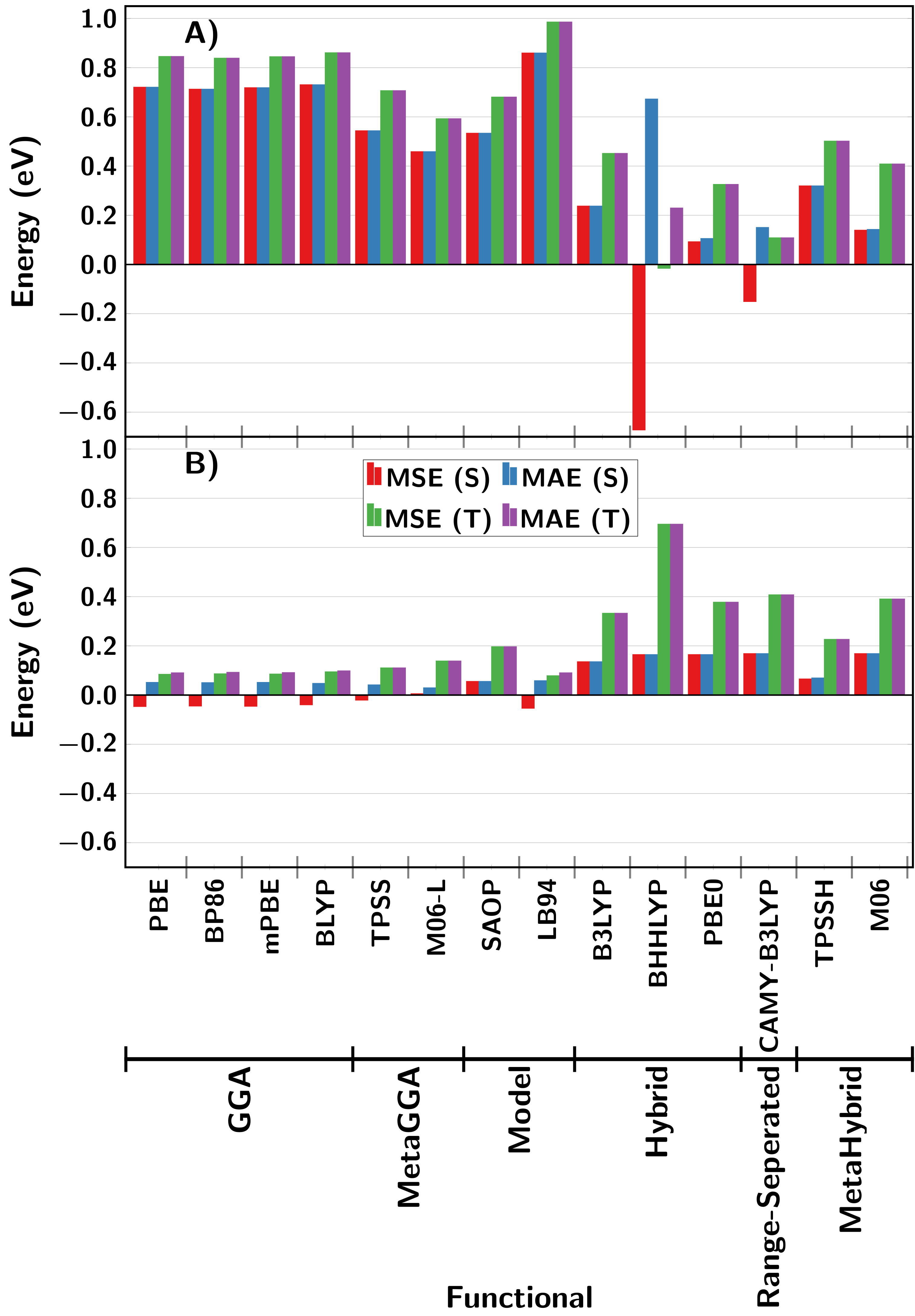}
\end{center}
\end{figure}

A simpler and more effective comparison for the 17 XC functionals considered  is to evaluate the mean signed errors (MSE) and the mean absolute errors (MAE) with respect to the MS-CASPT2(16,13) results.
These are plotted in Figure \ref{fig:excenrgfunct}A for the singlet and triplet states.
All the errors are reported for the first 8 singlet and 7 triplet excited states (i.e. excluding the T$_{1}$ and T$_{3}$).

One observation is that all the GGA functionals give errors in the vertical excitation energies which are almost identical.
Both singlet and triplet states excitation energies being underestimated but errors are typically smaller for the singlet states than for the triplet states, despite both the MSE and the MAE are of the same amplitude.
This is an indication that the errors are systematic and thus an empirical shift could be applied to align the results with experiment.
Similar observations hold for the meta-GGA and model XC functionals, albeit the errors are smaller with the exception of LB94.

On a side note, the geometry utilized for the TD-DFT of \rubpy\ was not identical to the structure used in the MS-CASPT2 calculation. Thus a comparison of the TD-DFT results using BP86 on the DFT optimized geometry and the RI-MP2 optimized geometry is shown in the Supporting Information (SI) Section \Romannumeral{1}. There, one sees that the fact that different structures which are employed in the TD-DFT calculations of \rubpy\ compared to the MS-CASPT2 does not invalidate any conclusions drawn, as the differences between the excited state energies was maximum 0.01 eV and for the oscillator strengths 0.001.

The hybrid XC functionals and their meta and range-separated counterparts show a general decrease in the size of the errors compared to MS-CASPT2(16,13).
The errors also appear systematic, except in the case of BHHLYP.
The unsystematic nature of the errors in BHHLYP may in part be attributed to the difficulties we had to achieve SCF convergence with this functional in this system, i.e. although convergence was achieved it might not correspond to the correct solution of the SCF.
Also as observed for the GGA functionals, the errors for the singlet states tend to be smaller than the errors for the triplet states.
However, the difference in the size of the MSE and MAE for singlets compared to triplets is larger for the hybrid XC functionals.

An initial conclusion to draw is that hybrid functionals show the best performance compared to the MS-CASPT2 results for the vertical excitation energies.
In particular PBE0 for the singlet excited states and CAMY-B3LYP for the triplet excited states seem to be very suitable functionals for describing vertical excitation energies in \rubpy.
This conclusion is in line with other studies, where hybrid functionals have shown superior performance in describing the electronic excited states of Cr(CO)$_{6}$ \cite{daniel_photochemistry_2015}, and many organic molecules \cite{silva-junior_benchmarks_2008,jacquemin_extensive_2009,jacquemin_assessment_2010}.

Despite the importance attributed  to reproduce vertical excitation energies, the energy gap between the excited states themselves is decisive in the fate of the system after irradiation, as the energetic gap between the states determines the probability for a radiationless transition.
With this motivation in mind, Figure \ref{fig:excenrgfunct}B displays the errors of the different functionals for the excited state energy gaps (as an example computed using the S$_{1}$ energy as a base line) against the MS-CASPT2(16,13) calculations  for 8 singlet and 7 triplet states.
Conspicuously, the trends on which XC functional perform best are now different to that obtained from just considering excitation energies from the ground state.
The energy gaps computed with the GGA XC functionals all involve an MSE and MAE of less than 0.1 eV for both the singlet and triplet states, with the singlet states having the lower error.
Two exceptions are  M06-L and SAOP, which underestimate the triplet gaps to a larger extent.

Within the model XC functionals, it is noticeable that the errors are smaller for LB94 than for SAOP when considering the excited state energy gaps, even though the opposite was true for the excitation energies.
The hybrid XC functionals predict singlet states that tend to have errors in their splitting of just over 0.1 eV.
The only exception is TPSSH which has an MAE of \textit{ca.}~0.07 eV.
However, the triplet state excited state energy gaps MSE and MAE are significantly larger than those observed for the GGA type XC functionals, typically 2 to 3 times larger.
The best performing hybrid functional is the TPSSH meta-hybrid functional which shows an MAE of 0.07 and 0.21 eV  for the singlets and triplets, respectively. Yet, the accuracy in energy gaps is not comparable to that obtained for the pure GGA functionals.

From these results alone one is left to conclude that the use of a standard GGA type XC functional, such as BP86 or PBE, should be more suitable to describe excited state dynamics of \rubpy than hybrid functionals.

\subsubsection{Basis set dependence}
Since the accuracy of TD-DFT is not always systematic with respect to the basis set, we analyze here its effect on the energies of the excited states of \rubpy\, taking as a reference the results obtained with BP86 and the TZ2P basis set, and employing the QZ4P, TZP and TZP(Ru)-DZP basis sets for comparison, all with no frozen core.
Any other parameter was kept identical.

\begin{figure}
\caption{Mean signed errors (MSE) and mean absolute errors (MAE) of the excitation energies with respect to $S_0$ (A) and the energy gaps with respect to $S_1$ for singlets (S) and triplet (T) states (B) of \rubpy\ computed with different basis sets against the BP86/TZ2P reference DFT calculation.}
 \label{fig:excenrgbasis}
\begin{center}
 \includegraphics[width=0.8\textwidth]{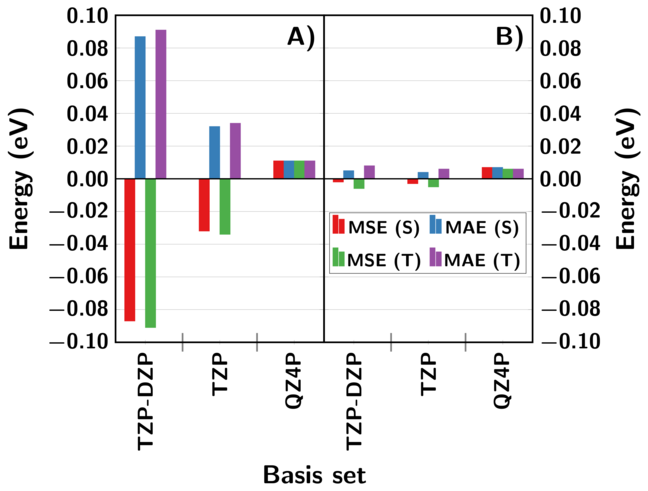}
\end{center}
\end{figure}

Figure \ref{fig:excenrgbasis}A shows the MAE and MSE for the singlets and triplet excitation energies relative to the S$_{0}$. Reducing the basis set size to the more economical TZP(Ru)-DZP combination only introduces a systematic error of 0.087 eV for the triplets and 0.091 eV for the singlets.
Using the larger TZP basis set on all atoms decreases the MAE and MSE by almost a factor of 2, while increasing to QZ4P has a minor effect on the MAE and MSE (around 0.01 eV), at the expense of a much larger computational time, especially, with respect to the TZ2P reference calculation.

The errors for the relative energies to the S$_{1}$ for the different basis sets are shown in panel \ref{fig:excenrgbasis}B.
Here one can see that all the basis sets tested have errors around 0.01 eV.
This indicates that the energetics of the excited states are relatively independent on the choice of basis set where the metal
uses a basis set of at least triple zeta quality.
Thus, computational effort is the primary deciding factor here.

We thus conclude that TZP on the ruthenium atom and a DZP basis set on all other atoms is a trade off between accuracy and computational effort.

\subsubsection{Solvation Dependence}
In this section, we consider the effect of solvation on the excited state energies of \rubpy.
Three different COSMO \cite{pye_implementation_1999,klamt_cosmo:_1993,klamt_conductor-like_1995,klamt_treatment_1996} solvent types were analyzed: COSMO1 is the infinite dielectric, COSMO2 is water and COSMO3 is acetonitrile.

\begin{figure}
\caption{ Mean signed errors (MSE) and mean absolute errors (MAE) of the excitation energies with respect to $S_0$ (A) and the energy gaps with respect to $S_1$ for singlets (S) and triplet (T) states (B) of \rubpy\ computed with different solvation models with respect to the BP86/TZ2P unsolvated reference calculation.}
\label{fig:excenrgsolvation}
\begin{center}
 \includegraphics[width=0.8\textwidth]{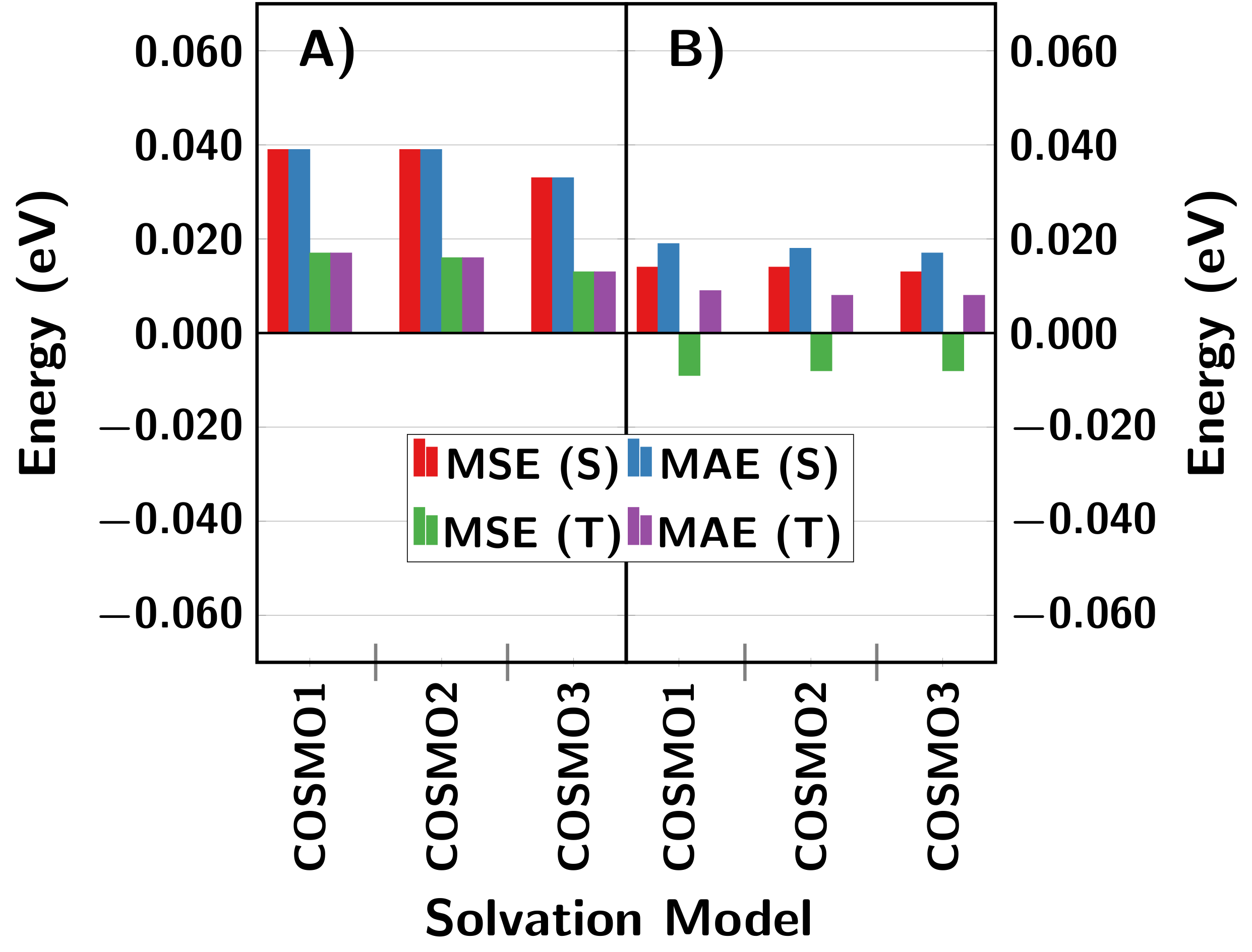}
\end{center}
\end{figure}

As in the previous section, we use BP86/TZ2P as a reference.
We compare the MAE and MSE errors of the energies of the excited states relative to the S$_{0}$ and of the energy gaps  relative to the S$_{1}$ of the unsolvated system calculated with BP86/TZ2P.
The results are shown in Figure \ref{fig:excenrgsolvation}A and \ref{fig:excenrgsolvation}B, respectively.
Interestingly, the effect of the solvent on the energy levels of the excited states is very small (maximum of 0.05 eV).
Also noticeable is that the solvent has a larger effect on the singlet states than the triplet states, except in one case (Figure \ref{fig:excenrgsolvation}A).
Within the COSMO models, the choice of parameters causes negligible differences up to $\sim$0.006 eV and the errors are systematic, as the MSE and MAE are of the same magnitude.
In any case, the MAE and MSE are small; therefore, from an energetic point of view the effect of solvation is rather small.

The errors in the energy gaps (Figure \ref{fig:excenrgsolvation}B) are essentially independent of the parameters of COSMO.
The changes in the dielectric constant lead to negligible energetic shifts (\textless 0.003 eV for the MAE).
However, the inclusion of solvent was shown to be a driver of of electron localization to one ligand in \rubpy\ \cite{moret_electron_2010} and so the inclusion of solvation can have a substantial effect, which can be seen dynamically but is not reflected in the energetics at the Franck-Condon geometry, at least
when using an implicit solvation method.

\color{black}
\subsubsection{Technical Settings Dependence}

For completeness, we also analyze the impact on the calculated results of different technical settings, such as the effect of TDA, the frozen core approximation as well as grid sizes and fitting types, as implemented in \adf.

TDA is recommended in cases where triplet states are involved~\cite{dinkelbach_assessment_2017} due to the possibility of triplet instabilities, particularly on the inclusion of HF exchange into the XC functional,
\cite{peach_influence_2011} or when exploring potential energy surfaces containing conical intersections.\cite{tapavicza_mixed_2008}
For \rubpy\ the errors are minimal, below 0.02 eV, with the error being largest for the singlet states, see Figure S1.


The effect of the frozen core approximation, different fitting methods, and grid sizes is minor and it is discussed in the SI, Section \Romannumeral{2}.
In conclusion, the most computationally efficient options, i.e., Becke Normal grid quality, ZLM spline normal grid quality and TDA are a good choice.

\subsubsection{State Character}
An optimum functional should not only provide the best excited state energies and energy gaps, but also preserve the state ordering and main character of the wavefunction compared to the benchmark.
As a common approximation, we assume that the CIS-type of wavefunctions derived from linear response TD-DFT can be interpreted as ordinary wavefunctions coming from ab initio theory.\cite{casida_time_1995}
Table \ref{tab:StaCharS} lists the orbital transitions with a weight greater than 0.1 that contribute to the state for both the MS-CASPT2 and the BP86/TZ2P TD-DFT (small frozen core) calculation of \rubpy.
The main orbitals involved in the transitions calculated with BP86 are displayed in the SI Figure S3 and for MS-CASPT2 in Figure \ref{fig:CASPT2orb}.
One can see that although the major character of the orbitals is the same, the DFT orbitals show more delocalisation in the occupied d orbitals and larger d character in the $\pi_2^{*} (e)$ and $\pi_3^{*} (e)$ orbitals.
This increase in delocalisation is not unexpected for GGA functionals\cite{mori-sanchez_many-electron_2006}.
Gratefully, not only the orbitals themselves but the main characters and ordering of the transitions calculated remain the same when using TD-BP86 compared to the MS-CASPT2(16,13).
This is the case until the S$_{5}$ as all the transitions are from or to an orbital belonging to the
the non-degenerate symmetry irreducible representations.
The S$_{6}$, S$_{7}$ and S$_{8}$ show larger variances between the BP86 and MS-CASPT2 calculations but the primary character is retained as all the transitions are mixes of the degenerate excitations from the d(e)$\rightarrow\pi^*$ but with
different mixing ratios between the two calculations. Also from Table \ref{tab:bipy-exc} one can see from the symmetry that the state ordering is the same for the singlets.

\begin{table}
\caption{Comparison of primary character of excitations within MS-CASPT2 and BP86 for the lowest 8 singlet excitations of \rubpy.}
\label{tab:StaCharS}
\begin{center}
\vspace{-0.7cm}
{\footnotesize
\begin{tabular}{ccccc}
\hline\hline
       \multicolumn{5}{c}{State Character}\\
 State& MS-CASPT2& Weight & BP86 & Weight \\
\hline
S$_1$&$d_{z^{2}}(a_1)\rightarrow\pi_1^*(a_2)$&0.90&$d_{z^{2}}(a_1)\rightarrow\pi_1^*(a_2)$&1.00\\[4pt]
S$_2$&$d_{z^{2}}(a_1)\rightarrow\pi_2^*(e)$&0.75&$d_{z^{2}}(a_1)\rightarrow \pi_2^*(e)$&0.99\\[4pt]
S$_3$&$d_{z^{2}}(a_1)\rightarrow\pi_3^*(e)$&0.81&$d_{z^{2}}(a_1)\rightarrow \pi_3^*(e)$&0.99\\[4pt]
\multirow{2}{*}{S$_4$}&$d_{xz}(e)\rightarrow\pi_1^*(a_2)$&0.75&$d_{xz}(e)\rightarrow \pi_1^*(a_2)$&0.74\\
&$d_{yz}(e)\rightarrow\pi_1^*(a_2)$&0.14&$d_{yz}(e)\rightarrow \pi_1^*(a_2)$&0.14\\[4pt]
\multirow{2}{*}{S$_5$}&$d_{yz}(e)\rightarrow\pi_1^*(a_2)$&0.70&$d_{yz}(e)\rightarrow \pi_1^*(a_2)$&0.74\\
&$d_{xz}(e)\rightarrow\pi_1^*(a_2)$&0.16&$d_{xz}(e)\rightarrow \pi_1^*(a_2)$&0.14\\[4pt]
\multirow{2}{*}{S$_6$}&$d_{yz}(e)\rightarrow\pi_2^*(e)$&0.70&$d_{xz} (e)\rightarrow \pi_3^*(e)$&0.47\\
&$d_{xz}(e)\rightarrow\pi_3^*(e)$&0.10&$d_{yz}(e)\rightarrow\pi_2^*(e)$&0.47\\[4pt]
\multirow{4}{*}{S$_7$}&$d_{xz}(e)\rightarrow\pi_3^*(e)$&0.40&$d_{yz}(e)\rightarrow \pi_2^*(e)$&0.30\\
&$d_{yz}(e)\rightarrow\pi_3^*(e)$&0.20&$d_{xz}(e)\rightarrow \pi_3^*(e)$&0.29\\
&$d_{xz}(e)\rightarrow\pi_2^*(e)$&0.17&$d_{xz}(e)\rightarrow \pi_2^*(e)$&0.13\\
&$d_{yz}(e)\rightarrow\pi_2^*(e)$&0.14&$d_{yz}(e)\rightarrow \pi_3^*(e)$&0.12\\[4pt]
\multirow{4}{*}{S$_8$}&$d_{xz}(e)\rightarrow\pi_3^*(e)$&0.38&$d_{xz}(e)\rightarrow \pi_2^*(e)$&0.30\\
&$d_{xz}(e)\rightarrow\pi_2^*(e)$&0.28&$d_{yz}(e)\rightarrow \pi_3^*(e)$&0.29\\
&$d_{yz}(e)\rightarrow\pi_3^*(e)$&0.24&$d_{yz}(e)\rightarrow \pi_2^*(e)$&0.13\\
&&&$d_{xz}(e)\rightarrow \pi_3^*(e)$&0.12\\
\cmidrule{1-5}
\multirow{2}{*}{T$_1$}&$d_{z^{2}}(a_1)\rightarrow\pi_1^*(a_2)$&0.74&$d_{z^{2}}(a_1)\rightarrow\pi_1^*(a_2)$&1.00\\
&$d_{z^{2}}(a_1)\rightarrow\pi_2^*(e)$&0.18&&\\[4pt]
T$_2$&$d_{z^{2}}(a_1)\rightarrow\pi_3^*(e)$&0.86&$d_{z^{2}}(a_1)\rightarrow \pi_2^*(e)$&1.00\\[4pt]
\multirow{2}{*}{T$_3$}&$d_{z^{2}}(a_1)\rightarrow\pi_2^*(e)$&0.73&$d_{z^{2}}(a_1)\rightarrow \pi_3^*(e)$&1.00\\
&$d_{z^{2}}(a_1)\rightarrow\pi_1^*(a_2)$&0.18&&\\[4pt]
\multirow{4}{*}{T$_4$}&$d_{yz}(e)\rightarrow\pi_1^*(a_2)$&0.30&$d_{xz}(e)\rightarrow \pi_1^*(a_2)$&0.97\\
&$d_{xz}(e)\rightarrow\pi_1^*(a_2)$&0.23&&\\
&$d_{xz}(e)\rightarrow\pi_2^*(e)$&0.19&&\\
&$d_{yz}(e)\rightarrow\pi_3^*(e)$&0.14&&\\[4pt]
\multirow{2}{*}{T$_5$}&$d_{yz}(e)\rightarrow\pi_1^*(a_2)$&0.54&$d_{yz}(e)\rightarrow \pi_1^*(a_2)$&0.97\\
&$d_{xz}(e)\rightarrow\pi_1^*(a_2)$&0.28&&\\[4pt]
\multirow{3}{*}{T$_6$}&$d_{xz}(e)\rightarrow\pi_2^*(e)$&0.40&$d_{xz}(e)\rightarrow \pi_2^*(e)$&0.47\\
&$d_{yz}(e)\rightarrow\pi_1^*(a_2)$&0.33&$d_{yz}(e)\rightarrow \pi_3^*(e)$&0.46\\
&$d_{yz}(e)\rightarrow\pi_3^*(e)$&0.12&&\\[4pt]
\multirow{4}{*}{T$_7$}&$d_{yz}(e)\rightarrow\pi_2^*(e)$&0.56&$d_{yz}(e)\rightarrow \pi_3^*(e)$&0.33\\
&$d_{xz}(e)\rightarrow\pi_3^*(e)$&0.12&$d_{xz}(e)\rightarrow \pi_2^*(e)$&0.32\\
&&&$d_{xz}(e)\rightarrow \pi_3^*(e)$&0.17\\
&&&$d_{yz}(e)\rightarrow \pi_3^*(e)$&0.17\\[4pt]
\multirow{4}{*}{T$_8$}&$d_{yz}(e)\rightarrow\pi_3^*(e)$&0.36&$d_{yz}(e)\rightarrow \pi_2^*(e)$&0.32\\
&$d_{yz}(e)\rightarrow\pi_2^*(e)$&0.32&$d_{xz}(e)\rightarrow \pi_3^*(e)$&0.32\\
&$d_{xz}(e)\rightarrow\pi_3^*(e)$&0.19&$d_{yz}(e)\rightarrow \pi_3^*(e)$&0.17\\
&&&$d_{xz}(e)\rightarrow \pi_2^*(e)$&0.17\\[4pt]
\multirow{3}{*}{T$_9$}&$d_{xz}(e)\rightarrow\pi_3^*(e)$&0.59&$d_{yz}(e)\rightarrow \pi_{2}^*(e)$&0.47\\
&$d_{yz}(e)\rightarrow\pi_3^*(e)$&0.21&$d_{xz}(e)\rightarrow \pi_3^*(e)$&0.47\\[4pt]
&$d_{xz}(e)\rightarrow\pi_2^*(e)$&0.13&&\\
\hline\hline
\end{tabular}}
\end{center}
\end{table}

The triplet states show more variation than the singlet states with respect to MS-CASPT2 showing a slight mixing of the T$_{1}$ and T$_{3}$, due to symmetry breaking.
Nevertheless, the major transition character of the T$_{1}$ to T$_{3}$ is reproduced by TD-BP86.
In TD-DFT the T$_{4}$ and T$_{5}$ are single transitions from the degenerate d orbitals the the LUMO $\pi_{1}^*$, but in MS-CASPT2 these states are strongly mixed. Again the primary transition character remains the same as TD-DFT with the major contributing transitions being from the degenerate d orbitals to the $\pi_1^*$. Furthermore, as was observed for the singlets, the T$_{7}$ to T$_{9}$ are all mixed transitions from the d(e) orbitals to the $\pi$(e) orbitals but with different ratios of mixing between the BP86 and MS-CASPT2 calculations.

\begin{figure}
\caption{Wave function overlap of the singlet (A) and triplet (B) states of \rubpy\ calculated using different density functionals with respect to BP86.}
 \label{fig:BPPBE}
\begin{center}
\includegraphics[width=\textwidth]{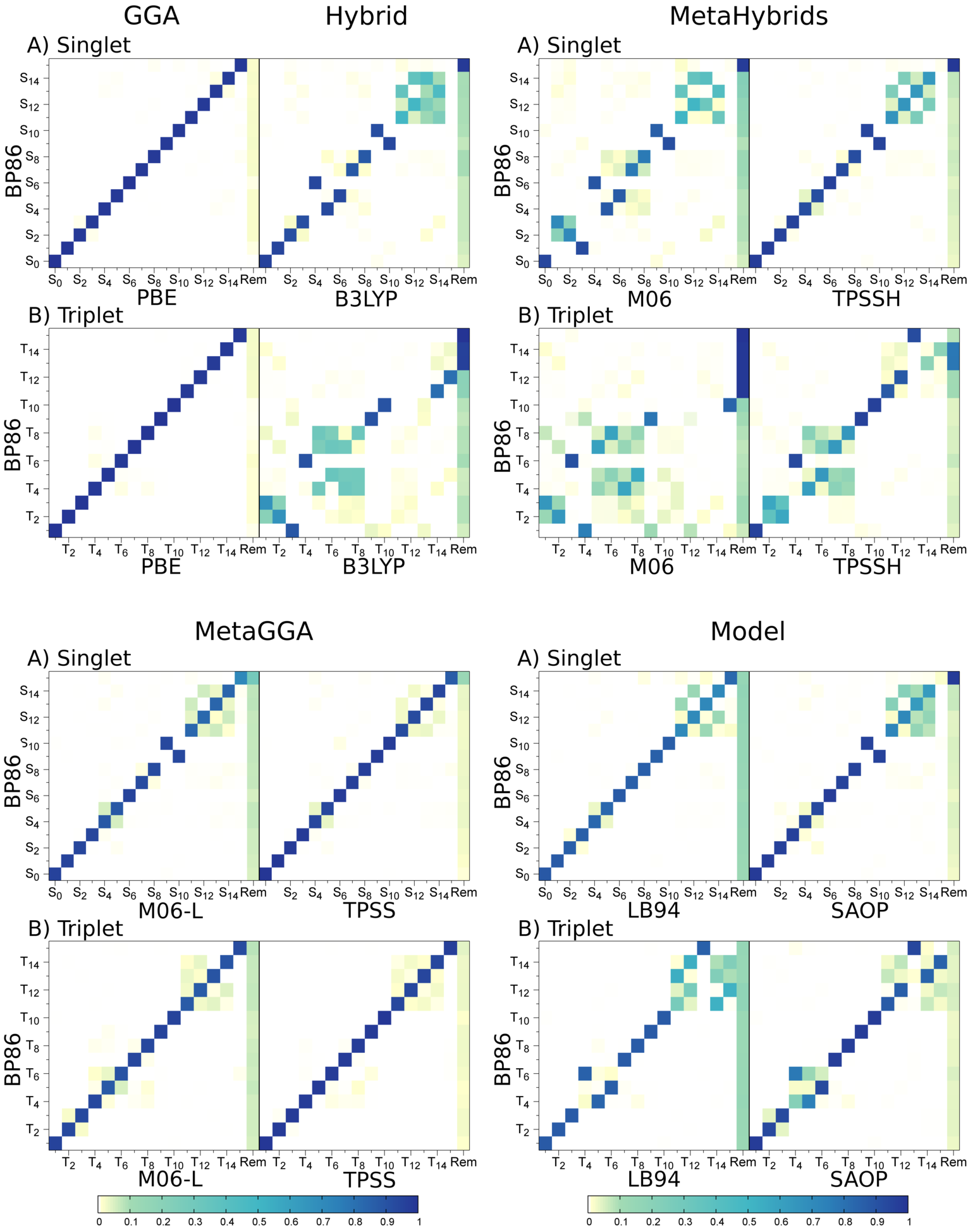}
\end{center}
\end{figure}

A comparison between BP86 and MS-CASPT2 in Table \ref{tab:bipy-exc} reveals  that the T$_7$, T$_8$ and T$_9$ have some slight reordering, with the T$_{9}$ from BP86 being the T$_{7}$ from MS-CASPT2.
However, from the overall perspective of the primary characters of each transition, BP86 reproduces the MS-CASPT2 calculation, even if the ratio of mixing of the different single orbital transitions is altered.
Since the BP86 functional using a TZ2P basis set provides the correct state ordering and character as compared to MS-CASPT2(16,13), it will be used as a baseline to compare the state characters provided by other functionals.
To this purpose, we present here a more efficient procedure than comparing visually all the transitions one-by-one will be employed.
Specifically, we employ the overlap between the two different wavefunctions, a method explained in Refs.~\citenum{plasser_efficient_2016,plasser_communication:_2016}, using a threshold of 0.99 for truncating the configuration vector.
We compare wavefunctions obtained with the same technical details across all the methods, except in the meta-functionals, where the overlaps were calculated with respect to BP86/TZ2P  using no frozen core approximation as it is not possible to include frozen core orbitals for meta-functionals.
Figures \ref{fig:BPPBE} shows the wavefunction overlaps of the excited states of \rubpy\ calculated using PBE and B3LYP, using as a reference the BP86 calculation.
Each state of the new functional is projected onto the reference BP86 one and then plotted with regards to the percentage of overlap between the states.\cite{plasser_communication:_2016}
The beauty of these diagrams is that it is straightforward to see any change. For instance, changing from PB86 to the PBE functional has negligible effects on the state characters since the primary character as well as the state ordering is retained for both the singlets and triplets states and thus no off-diagonal elements are present, see Figure~\ref{fig:BPPBE}.
This is also true for all the GGA functionals tested here (see Figure S6 of SI).

Larger differences than for GGA functionals are encountered for hybrid functionals, for B3LYP and other hybrid functionals (Figure S7 of SI).
For the singlet states, the state character obtained with BP86 is retained up to the S$_{10}$ with only minimal contributions from other states, while more significant mixing is found after S$_{10}$.
Additionally, the highest state calculated with BP86, S$_{15}$, is not present in the 15 excitations obtained from the B3LYP calculation.
More importantly, the state ordering of the states which retain single state character is also modified with B3LYP.
As the BP86 state ordering compares well with the reference MS-CASPT2, which has good agreement with experiment, we conclude that the change of state ordering by B3LYP is incorrect.
B3LYP predicts even a  more pronounced mixing of state character for the triplet states.
The state character predicted by BP86  is only preserved in six states, while many other states show a strong mixing of excitations, in comparison to MS-CASPT2.
Moreover, the three highest triplet excitations calculated with BP86 are not found in the B3LYP excitation space.

The wavefunction overlaps of meta-hybrid functionals tested (Figure~\ref{fig:BPPBE}), illustrate that such occurrences in the state character mixing and change of state ordering are also commonplace.
The calculated overlaps for a few meta-GGA and model functionals suggest that these functionals give very similar results as to the GGA functionals for the state characters.

\begin{figure}
\caption{Wave function overlap of the singlet (A) and triplet (B) states of \rubpy\ calculated using Different implicit solvents (COSMO1, COSMO2, COSMO3) with respect to the unsolvated reference calculation (BP86)}
 \label{fig:solv}
\begin{center}
 \includegraphics[width=0.9\textwidth]{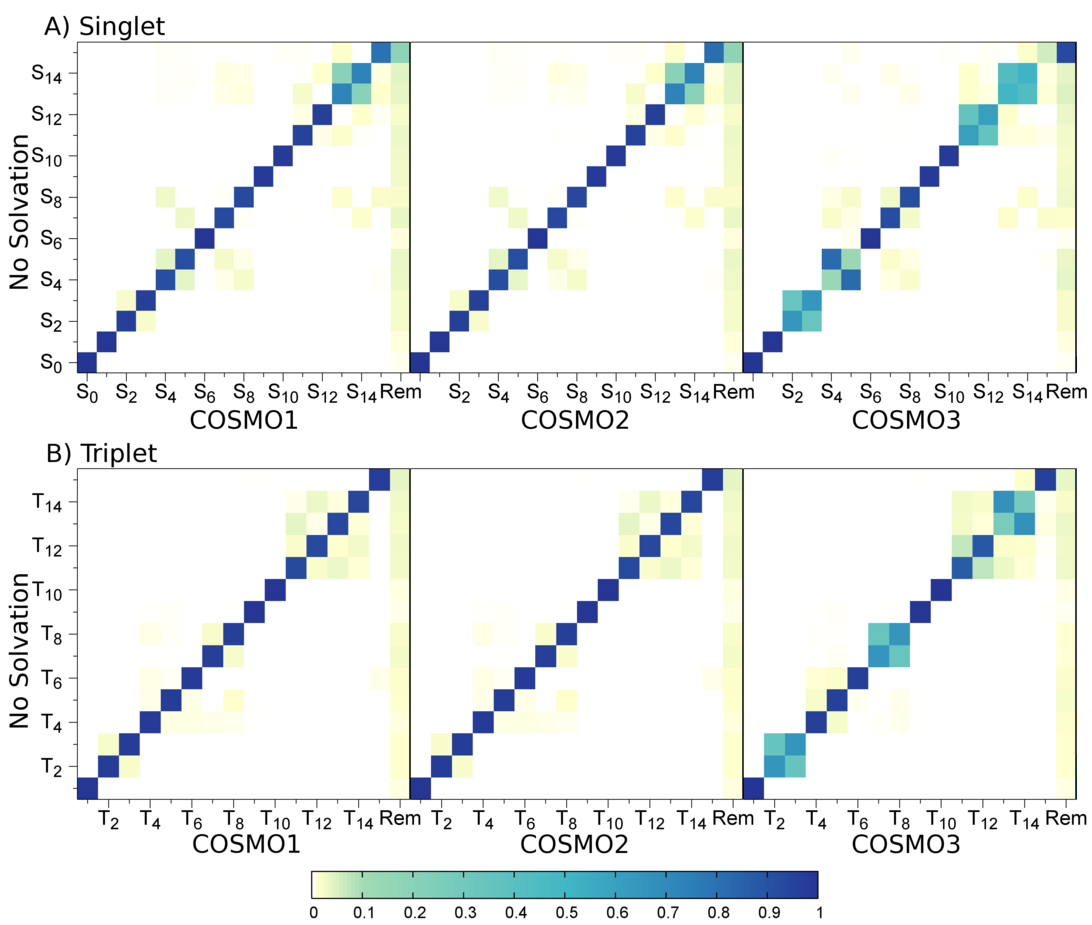}
\end{center}
\end{figure}

The changes in the state characters using different solvation models with BP86 are shown in Figure~\ref{fig:solv}.
What one can observe for the implicit solvation models, COSMO1 and COSMO2, is that the state characters are well preserved from the unsolvated calculations.
However, there is a slight increase in both the mixing and the orthogonal complement of the highest 3 singlet states.
COSMO3 shows more mixing of the degenerate excited states and the BP86 S$_{15}$ is only marginally represented in the calculated COMSO3 excitation space.
The triplets show the same behaviour for the mixing of the degenerate states, but all the BP86 triplet states are accounted for within the COSMO3 excitation window.
The effects of solvation on B3LYP can be found in Figure S8 of the SI.

Finally, concerning the technical details (Figures S9 and S10), one sees that reducing the size of the Becke grid from good to normal induces a strong mixing of the degenerate S/T$_{2}$ and S/T$_{3}$ and to a smaller degree the other degenerate excited states.
The fitting options for the electron density have no significant effect on either the state ordering or character and the ZLM normal grid is sufficient, as indicated by the excitation energy error comparison.
The use of TDA (Figure S11) has also no effect on the excited triplet states but the S$_9$ has swapped with  S$_{7}$ with TDA, what is not particularly severe since all S$_7$, S$_{8}$ and S$_{9}$ states are almost degenerate (0.03 eV difference) in the reference calculations.

\subsection{\rupy}
This section discusses the effect of the different density functionals on the energies and state characters of \rupy\ in comparison to the results obtained for the MS-CASPT2 benchmark calculation.

\subsubsection{Density Functional Dependence}
\label{sec:DFTdep2}
In Table \ref{tab:py-exc} we collect the spin-free vertical excitation energies and oscillator strengths of \rupy obtained with MS-CASPT2(16,13), and as an example those obtained with BP86 and B3LYP using the TDA approximation.
As previously seen for \rubpy, BP86 underestimates the energies more severely than B3LYP, 0.35 eV and 0.04 eV for the S$_{1}$ and 0.44 eV and 0.46 eV for the T$_{1}$, respectively compared to MS-CASPT2(16,13).

\def\mr#1{\multirow{-2}{*}[1pt]{$\left.\rule{0cm}{2.5ex}\right\}$#1}}
\begin{table}[h]
\caption{Symmetries, characters, spin-free MS-CASPT2(16,13), BP86 and B3LYP with TDA excitation energies
(in eV) and oscillator strengths $f$ of the lowest four singlet and triplet excited states of \rupy.}
\label{tab:py-exc}
\centering
\begin{tabular}{ccccccc}
\hline\hline
&\multicolumn{2}{c}{MS-CASPT2(16,13)}&\multicolumn{2}{c}{BP86 TD-DFT}&\multicolumn{2}{c}{B3LYP TD-DFT}\\
     State  & $\Delta E$ & $f$& $\Delta E$ & $f$ & $\Delta E$ & $f$ \\
     \midrule
  $S_1$  & 3.03 & 0.002&2.78&0.005&2.99&0.002 \\
  $S_2$ & 3.03 & 0.002&2.78&0.005&2.99&0.002 \\
  $S_3$  & 3.66 & 0.000&2.88&0.000&3.05&0.000 \\
  $S_4$ & 3.79 & 0.000&2.95&0.000&3.20&0.0000 \\
     \cmidrule(rl){1-7}
      $T_1$ &2.92  &--&2.48&--&2.46&--\\
      $T_2$ &2.93  &--&2.48&--&2.68&--\\
      $T_3$ &3.09 &--&2.56&--&2.68&--\\
      $T_4$ &3.11 &--&2.70&--&2.72&--\\
\hline\hline
\end{tabular}
\end{table}

\begin{figure}
\caption{Mean signed errors (MSE) and mean absolute errors (MAE) of the excitation energies (without TDA) (panel A) and the energy gaps for singlets (S) and triplet (T) states (panel B) of \rupy\ computed against MS-CASPT2(16,13). Note that the scale is different from that used in \rubpy.}
 \label{fig:excenrgfunctpy}
\begin{center}
 \includegraphics[width=\textwidth]{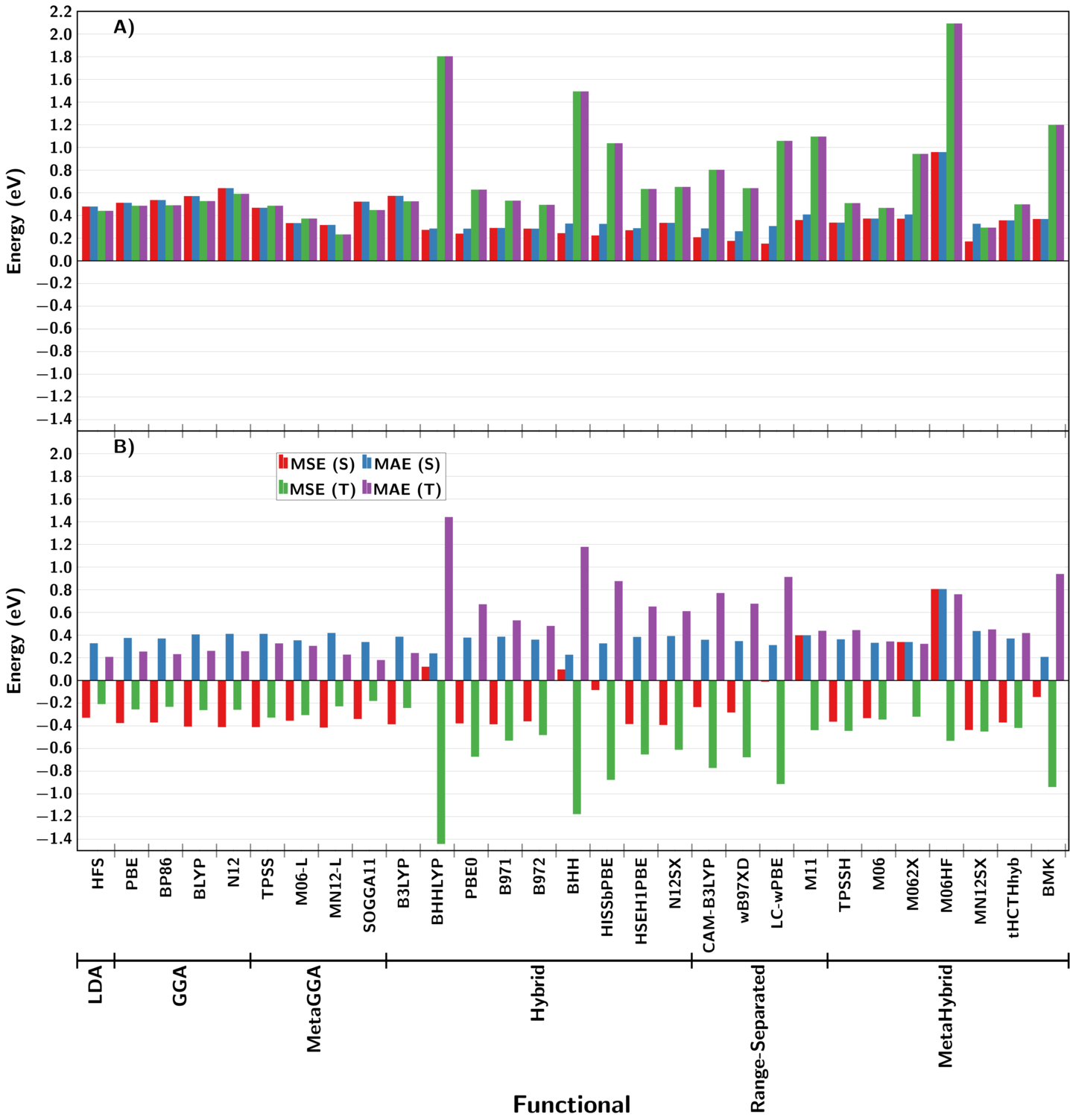}
\end{center}
\end{figure}

\begin{figure}
\caption{Mean signed errors (MSE) and mean absolute errors (MAE) of the excitation energies (with TDA) (panel A) and the energy gaps for singlets (S) and triplet (T) states (panel B) of \rupy\ computed against MS-CASPT2(16,13). Note that the scale is different from that used in Figure~\ref{fig:excenrgfunctpy} .}
 \label{fig:excenrgfuncttdapy}
\begin{center}
 \includegraphics[width=\textwidth]{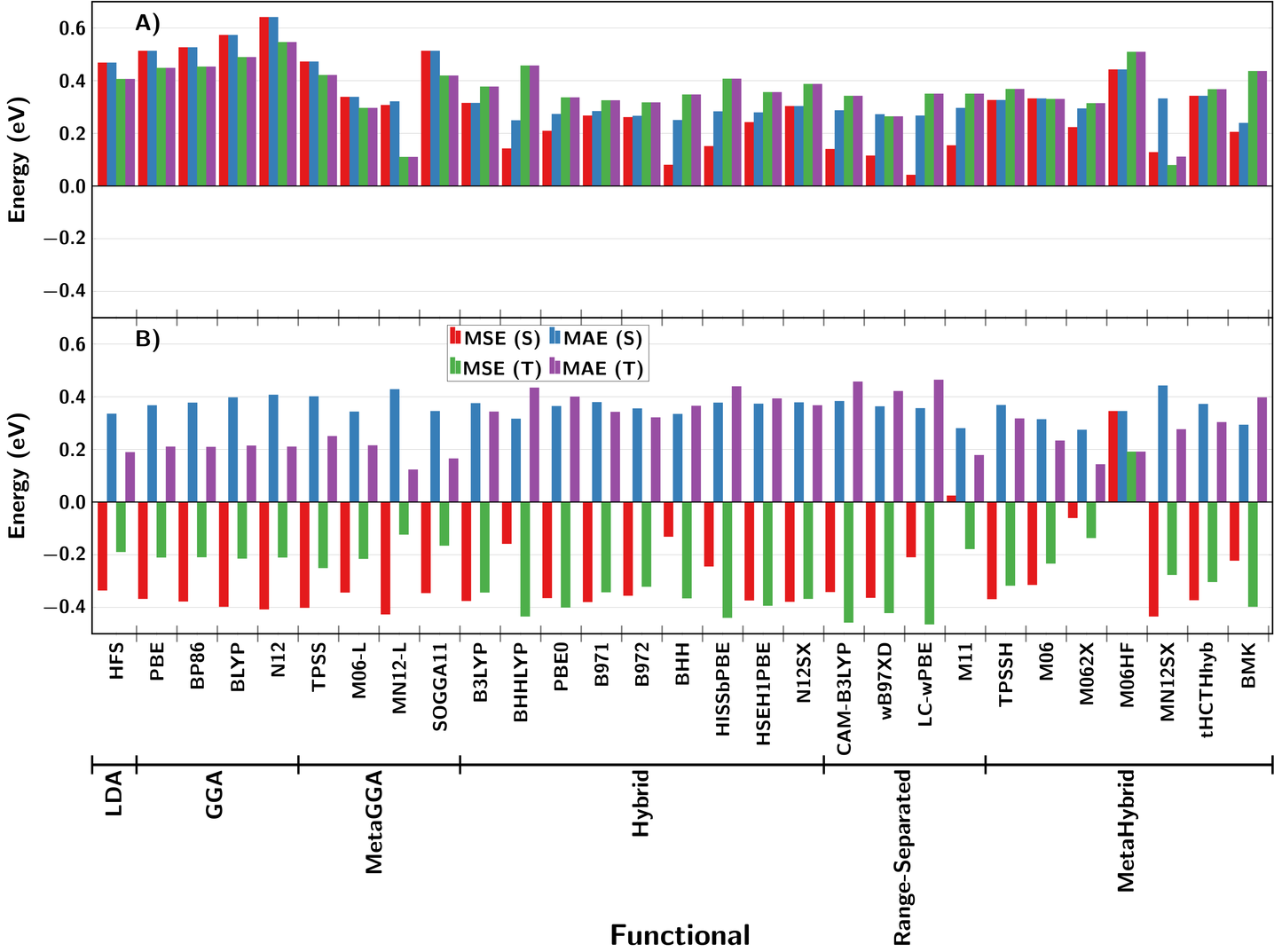}
\end{center}
\end{figure}

For all the 29 XC functionals employed, the MSE and MAE with respect to MS-CASPT2(16,13) vertical excitation energies for the first 4 singlet and 4 triplet excited states is presented in Figures \ref{fig:excenrgfunctpy}A and \ref{fig:excenrgfuncttdapy}A without and with TDA, respectively.
Regardless of TDA, the LDA and GGA XC functionals provide similar MSE and MAE values underestimating to a similar extent both the singlet and triplet energies, within 0.1 eV difference.
Important is that the MSE and MAE are of the same size, indicating that the error in the excitation energies is systematic.
The use of meta- variants of the GGA functionals tends to decrease the errors in the vertical transition energies, with the exception of SOGGA11.

The hybrid XC functionals and their meta and range-separated counterparts show more variation in their MSE and MAE values compared to the GGA and LDA XC functionals. Nevertheless, 13 out of the 20 hybrid XC functionals investigated show smaller errors for the singlet vertical transition energies than the GGA functionals, whilst several have errors on par with those observed for the GGA XC functionals.
M06HF shows significantly increased errors for the singlet energies.
For the triplet vertical excitation energies 12 out of the 20 hybrid XC functionals are better than GGA functionals.

Interestingly, the commonly used B3LYP functional here shows an average performance almost identical to the GGA functionals, although a detailed comparison of B3LYP with BP86 shows differences for the different states. In general, in the hybrid functionals the errors can be seen as systematic, but there are exceptions in particular functionals.

In contrast to \rubpy, here the use of TDA has a noticeable effect on the triplet states calculated using the hybrid functionals. For the LDA and GGA XC functionals one observes a minor reduction in the errors for the triplet states. For the hybrid XC functionals the errors for the triplet states is now similar or or smaller than for the GGA XC functionals. This is most notable in the case of M06HF where both the singlet and triplet state MSE and MAE are reduced by at least twofold. An important note is also that several of the functionals that showed more systematic errors without TDA now have unsystematic errors on the inclusion of TDA. An example here is MN12SX for the triplet excited states and BMK for the singlet excited states.

An initial conclusion from this data would be to use a hybrid XC functional with TDA for the calculation of the vertical excitation energies of \rupy, similar to the initial conclusion drawn for \rubpy.
In this case, B972 for both the singlet and triplet excited states seems to be a suitable functional.

The MSE and MAE for the energy gaps of the singlet and triplet excited states relative to the S$_{1}$
are shown in Figures \ref{fig:excenrgfunctpy}B and \ref{fig:excenrgfuncttdapy}B without and with TDA, respectively.
Unlike the observation for \rubpy\ the trend of GGA XC functionals providing better relative energies than the hybrid XC functionals when not using TDA is not so well defined. The reduction in errors going from the vertical excitation energies to the relative energies is not observed for the singlet energy gaps calculated with GGA XC functionals, although the MSE and MAE for the triplets is reduced.
However, the hybrid XC functionals without TDA on average show larger errors in their energy gaps relative to the S$_{1}$ than the GGA XC functionals, specifically for the triplet energy gaps. There are a few hybrid XC functionals which provide similar MSE and MAE  for the energy gaps as to the GGA XC functionals, namely B3LYP, M06 and M062X.

As observed for the vertical excitation energies, the use of TDA has a larger effect on the MSE and MAE of the hybrid XC functionals than the GGA and LDA XC functionals. Though the errors are marginally reduded for triplet energy gaps by using TDA with the GGA XC functionals, the change is negligible. However, the trend observed that the GGA XC functionals give smaller errors for the energy gaps compared to the hybrids does not hold for \rupy\ as all the tested functionals provide similar MSE and MAE for the singlet excited states. Nonetheless, for the triplet excited states the GGA XC functionals still hold an advantage over the hybrid XC functionals, which give an MSE and MAE approximately twice that of the GGA XC functionals.

Furthermore, the errors calculated for the GGA functionals are systematic in nature, which does not hold true for all the calculated hybrid functionals. From these results, same as for \rubpy, one may conclude that the use of standard GGA functionals, such as PBE or BP86, should be more suitable to describe energy gaps than the hybrid XC functionals, even though several of the hybrid XC functionals may also be suitable.

\subsubsection{State Character}
As in \rubpy, BP86, but in this case with TDA (BP86/TDA) as it energetically showed adequate performance, will be used to compare the performance of the other functionals.
Table \ref{tab:StaCharSpy} lists the major orbital transitions obtained with MS-CASPT2 and BP86/TDA.
The corresponding orbitals are shown in Figure~\ref{fig:CASPT2orb} and Figure S4, respectively.
The major character of the BP86/TDA orbitals coincides with those from MS-CASPT2 although the DFT orbitals are more delocalised  with more inclusion of $\pi$ orbitals from the pyridine ligands.

When comparing the character of the transitions one sees that the S$_{1}$ and S$_{2}$ of MS-CASPT2 are reproduced by BP86/TDA.
Both states are composed of single transitions from the HOMO $d_{xy}$ to the unoccupied antibonding orbitals with the NO ligand.
However, larger deviations occur for the S$_{3}$ and S$_{4}$ where both have a significant contribution of the $d_{xy}$ to the antibonding d$_{x^2-y^2}$ orbital.
In general, the transitions predicted by BP86/TDA are in qualitative agreement with the MS-CASPT2 calculation, even if parts of the S$_{3}$ and S$_{4}$ states are not reproduced.

The triplet states show the same behaviour as the singlet states, with both the T$_{1}$ and T$_{2}$ of the MS-CASPT2 calculation being reproduced by BP86/TDA.
Moreover, the T$_{3}$ of MS-CASPT2 corresponds to the T$_{3}$ calculated with BP86/TDA for the major contributors to the excited state.
However, BP86/TDA shows more transitions mixing in and so a lower coefficient for the main transitions.
The T$_{4}$ has no correspondence between the MS-CASPT2 and BP86/TDA calculation.

\begin{table}
\caption{Comparison of primary character of excitations  within MS-CASPT2 and BP86 with TDA (BP86/TDA) for the lowest four singlet and triplet excitations of \rupy.}
\label{tab:StaCharSpy}
\begin{center}
\begin{tabular}{ccccc}
\hline\hline
       \multicolumn{5}{c}{State Character}\\
 State& MS-CASPT2& Weight & BP86/TDA & Weight \\
\hline
\multirow{1}{*}{S$_1$}&$d_{xy}\rightarrow d_{yz}-\pi_y^*$&0.85&$d_{xy}\rightarrow d_{xz}-\pi_{x}^{*}$&0.99\\[4pt]
\multirow{1}{*}{S$_2$}&$d_{xy}\rightarrow d_{xz}-\pi_x^{*}$&0.85&$d_{xy}\rightarrow d_{yz}-\pi_{y}^{*}$&0.99\\[4pt]
\multirow{4}{*}{S$_3$}&$d_{xy}\rightarrow d_{x^2-y^2}-\sigma_{Py}$&0.41&$d_{xz}+\pi_{x}^{*}\rightarrow d_{yz}-\pi_{y}^{*}$&0.28\\
&$d_{xz}+\pi_{x}^{*}\rightarrow d_{yz}-\pi_{y}^{*}$&0.21&$d_{yz}+\pi_{y}^{*}\rightarrow d_{xz}-\pi_{x}^{*}$&0.28\\
&$d_{yz}+\pi_{y}^{*}\rightarrow d_{xz}-\pi_{x}^{*}$&0.17&$d_{xz}+\pi_{x}^{*}\rightarrow d_{xz}-\pi_{x}^{*}$&0.22\\
&&&$d_{yz}+\pi_{y}\rightarrow d_{yz}-\pi_{y}^{*}$&0.22\\[4pt]
\multirow{3}{*}{S$_4$}&$d_{yz}+\pi_{y}^{*}\rightarrow d_{xz}-\pi_{x}^{*}$&0.32&$d_{xz}+\pi_{x}^{*}\rightarrow d_{xz}-\pi_{x}^{*}$&0.48\\
&$d_{xy}\rightarrow d_{x^2-y^2}-\sigma_{Py}$&0.31&$d_{yz}+\pi_{y}^{*}\rightarrow d_{yz}-\pi_{y}^{*}$&0.48\\
&$d_{xz}+\pi_{x}^{*}\rightarrow d_{yz}-\pi_{y}^{*}$&0.18&&\\
\cmidrule{1-5}
\multirow{1}{*}{T$_1$}&$d_{xy}\rightarrow d_{yz}-\pi_y^*$&0.88&$d_{xy}\rightarrow d_{xz}-\pi_{x}^{*}$&0.99\\[4pt]
\multirow{1}{*}{T$_2$}&$d_{xy}\rightarrow d_{xz}-\pi_x^{*}$&0.88&$d_{xy}\rightarrow d_{yz}-\pi_{y}^{*}$&0.99\\[4pt]
\multirow{4}{*}{T$_3$}&$d_{yz}+\pi_{y}^{*}\rightarrow d_{yz}-\pi_{y}^{*}$&0.45&$d_{yz}+\pi_{y}\rightarrow d_{yz}-\pi_{y}^{*}$&0.22\\
&$d_{xz}+\pi_{x}^{*}\rightarrow d_{xz}-\pi_x^{*}$&0.45&$d_{xz}+\pi_{x}\rightarrow d_{xz}-\pi_{x}^{*}$&0.22\\
&&&$d_{yz}+\pi_{y}\rightarrow d_{xz}-\pi_{x}^{*}$&0.19\\
&&&$d_{xz}+\pi_{x}\rightarrow d_{yz}-\pi_{y}^{*}$&0.19\\[4pt]
\multirow{2}{*}{T$_4$}&$d_{xy}\rightarrow d_{x^2-y^2}-\sigma_{Py}$&0.67&$d_{yz}+\pi_{y}\rightarrow d_{xz}-\pi_{x}^{*}$&0.42\\
&&&$d_{xz}+\pi_{x}\rightarrow d_{yz}-\pi_{y}^{*}$&0.42\\
\hline\hline
\end{tabular}
 \end{center}
\end{table}



\begin{figure}
\caption{Wave function overlap of the singlet (A) and triplet (B) states of \rupy\
calculated using different functionals with respect to BP86/TDA.}
 \label{fig:py_overl}
\begin{center}
 \includegraphics[width=\textwidth]{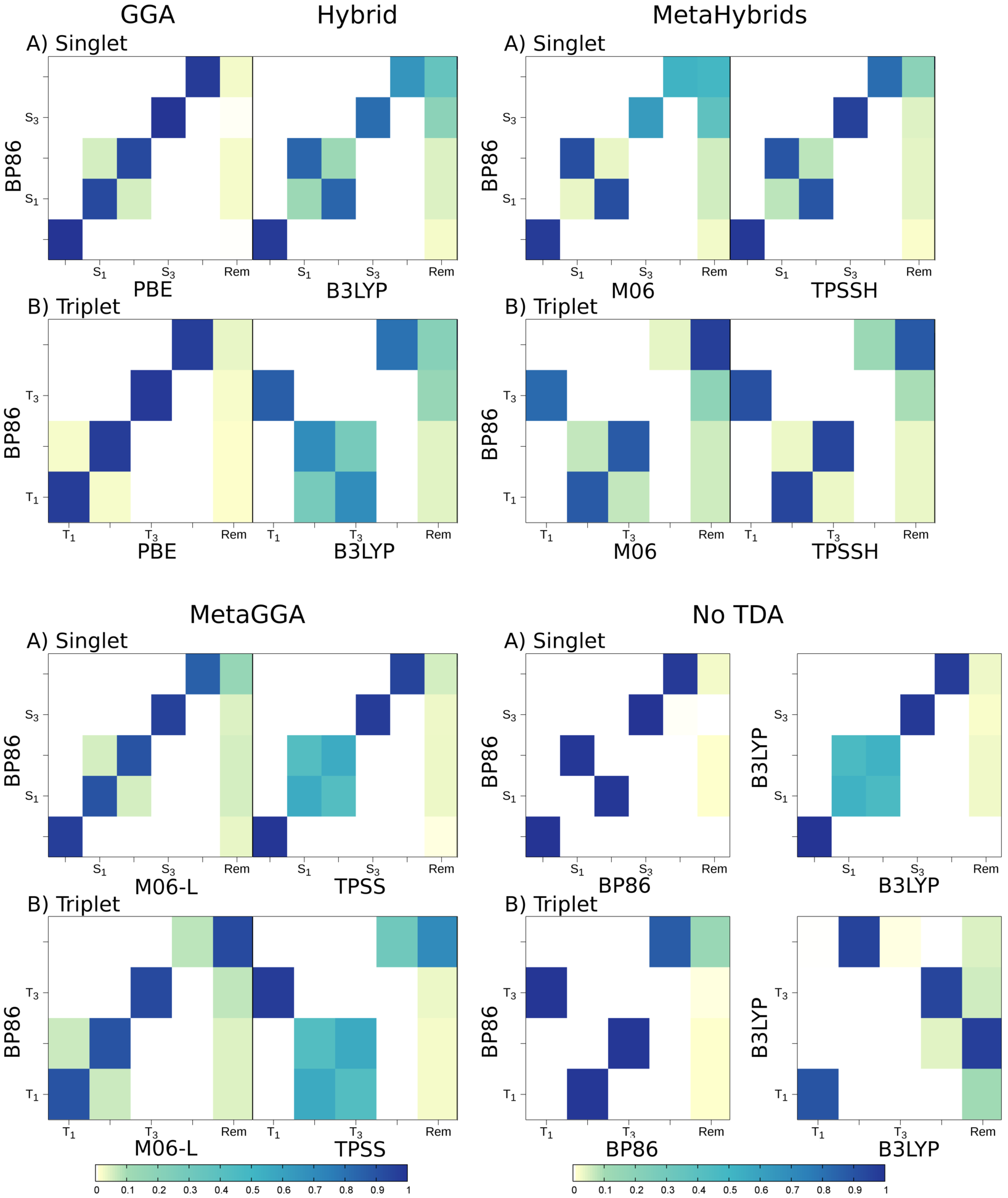}
\end{center}
\end{figure}

Since the BP86/TDA functional reproduces qualitatively the singlet excited states in the correct order and lowest three triplet states compared to MS-CASPT2, we take BP86/TDA as a reference to assess the performance of the other functionals in describing the  excited states of \rupy\ using the overlap between the two different wave functions.
These are compared with the same technical details and including the TDA.
Due to the number of functionals, only a few are selected as representatives of different functional categories (only the range-separated do not have a representative shown here).
Figure \ref{fig:py_overl} shows the wavefunction overlaps for selected functionals for \rupy\ as well as a comparison of the effect of TDA on the wave function for BP86 and B3LYP.
It is now very easy to see that PBE gives very similar state characters as BP86 with a slight mixing of the S$_{1}$ and S$_{2}$ states; the same mixing is observed for the  T$_{1}$ and T$_{2}$  triplet states. These observations hold for all the GGA functionals tested here, with N12 showing stronger mixing of the two lowest excited states in both singlets and triplets.

In the case of the hybrid functional B3LYP one observes that the S$_{3}$ and S$_{4}$ have slightly different state characters than BP86/TDA.
Much larger differences are present for the triplet states. Now the T$_{1}$ and T$_{2}$ have switched places with the T$_{3}$ and the T$_{1}$ and T$_{2}$ have also switched, but as with PBE the T$_{1}$ and T$_{2}$ show mixing compared to the BP86/TDA triplet states.
The other hybrid functionals tested (wave function overlaps not shown) also show varying degrees of state mixing and state reordering in both the singlet and triplet states as well as completely differing state characters of the lower excited states in some cases (i.e., BHHLYP).
Since the state ordering of the lowest three triplet states is reproduced well by the BP86/TDA calculation compared to the MS-CASPT2, one can conclude that the state reordering predicted by B3LYP for the triplets is incorrect.

The chosen metaHybrid XC functionals (M06 and TPSSH) show the same  behaviour as B3LYP for both the singlet and the triplet excited states. However, M06 shows a larger
deviation of the reproduced state character of the S$_{3}$ and S$_{4}$  of the reference BP86/TDA calculation compared to B3LYP or TPSSH. For the triplet states the lowest 3 triplets have been reordered as was the case for B3LYP and the fourth tiplet state in both M06 and TPSSH for the majority does not correspond to the fourth triplet excited state calculated with BP86/TDA. Thus the same conclusion as for B3LYP must be derived.

The calculated overlaps for two metaGGA functionals (M06-L and TPSS) provide for the singlet excited states the same picture as PBE, although TPSS shows more state mixing of the S$_{1}$ and S$_{2}$ than M06-L.
For the triplet states TPSS predicts an incorrect reordering of the lowest three triplet states, while  M06-L reproduces the state characters of the lowest 3 triplets and as with PBE the T$_{1}$ and T$_{2}$ show some mixing. The only deviation is for the T$_{4}$ which in M06-L only retains a small margin of the state character calculated with BP86/TDA.

Finally, we find it interesting to consider the effect of TDA on the state characters.  Opposite to what it was observed for \rubpy\, the effect of TDA in \rupy\ is more pronounced.
In the singlet excited states of BP86 the S$_1$ and S$_2$ are switched. Even larger deviations are observed for the triplet states: the T$_{3}$ is the T$_{1}$ when not using TDA and the T$_{1}$ and T$_{2}$ become the T$_2$ and T$_3$.
With B3LYP, again the singlet states are mildly affected  with the S$_{1}$ and S$_{2}$ being more mixed when not using TDA but the triplets are severely affected by the inclusion of TDA.
The T$_{1}$ remains primarily the same but the T$_{2}$ calculated when including TDA is not present at all in the absence of TDA and the T$_{3}$ and T$_{4}$ are reordered.
Thus, one can conclude that the use of TDA is highly recommended for reproducing the correct state ordering in \rupy, especially for the triplet excited states.

\subsection{\rupapy}
The following discussion includes the effect of the different density functionals on the energies and state characters of \rupapy\ in comparison to the results obtained for the MS-CASPT2 benchmark calculation.


\subsubsection{Density Functional Dependence}
\label{sec:DFTdep3}
The results of MS-CASPT2(18,14), taken from Ref. \citenum{freitag_theoretical_2014} and those obtained with BP86/TDA and B3LYP/TDA are listed in Table ~\ref{tab:papy-exci-dft}.
A comparison of the MS-CSAPT2 and BP86 without TDA is also made in Ref.~\citenum{freitag_theoretical_2014}.

As was seen for \rubpy, BP86 underestimates the energies more severely than B3LYP, 0.74 eV and 0.51 eV for the S$_{1}$ and 0.58 eV and 0.61 eV for the T$_{1}$, respectively compared to MS-CASPT2(18,14).

\begin{table}[h]
\caption{Symmetries, characters, spin-free MS-CASPT2(18,14)$^a$, BP86 and B3LYP excitation energies with TDA (in eV) and oscillator strengths $f$ of the lowest singlet and triplet
excited states of \rupapy.}
\label{tab:papy-exci-dft}
\begin{center}
\begin{tabular}{ccccccc}
\hline\hline
&\multicolumn{2}{c}{MS-CASPT2(18,14)}&\multicolumn{2}{c}{BP86/TDA}&\multicolumn{2}{c}{B3LYP/TDA}\\
     State     & $\Delta E$ & $f$  & $\Delta E$ & $f$ & $\Delta E$ & $f$\\
     \midrule
     $S_1$ &2.83&0.0000&2.09&0.000&2.32&0.000\\
     $S_2$ &3.11&0.0090&2.26&0.000&2.84&0.030\\
     $S_3$ &3.22&0.0003&2.30&0.001&3.06&0.003\\
  $S_4$ &3.50&0.0250&2.78&0.033&3.11&0.001\\
     \cmidrule(rl){1-7}
      $T_1$ &2.46&--&1.88&--&1.85&--\\
      $T_2$ &2.55&--&1.98&--&2.10&--\\
      $T_3$ &3.02&--&2.14&--&2.79&--\\
      $T_4$ &3.09&--&2.23&--&2.80&--\\
      $T_5$ &3.13&--&2.59&--&2.85&--\\
      $T_6$ &3.54&--&2.66&--&2.90&--\\
\hline\hline
\end{tabular}\\
 $^a$ Characters for MS-CASPT2 calculation from Ref.~\citenum{freitag_theoretical_2014}
\end{center}
\end{table}

\begin{figure}
\caption{Mean signed errors (MSE) and mean absolute errors (MAE) of the excitation energies computed without TDA (panel A) and the energy gaps for singlets (S) and triplet (T) states (panel B) of \rupapy\ computed against MS-CASPT2(18,14). }
 \label{fig:excenrgfunctpapy}
\begin{center}
 \includegraphics[width=\textwidth]{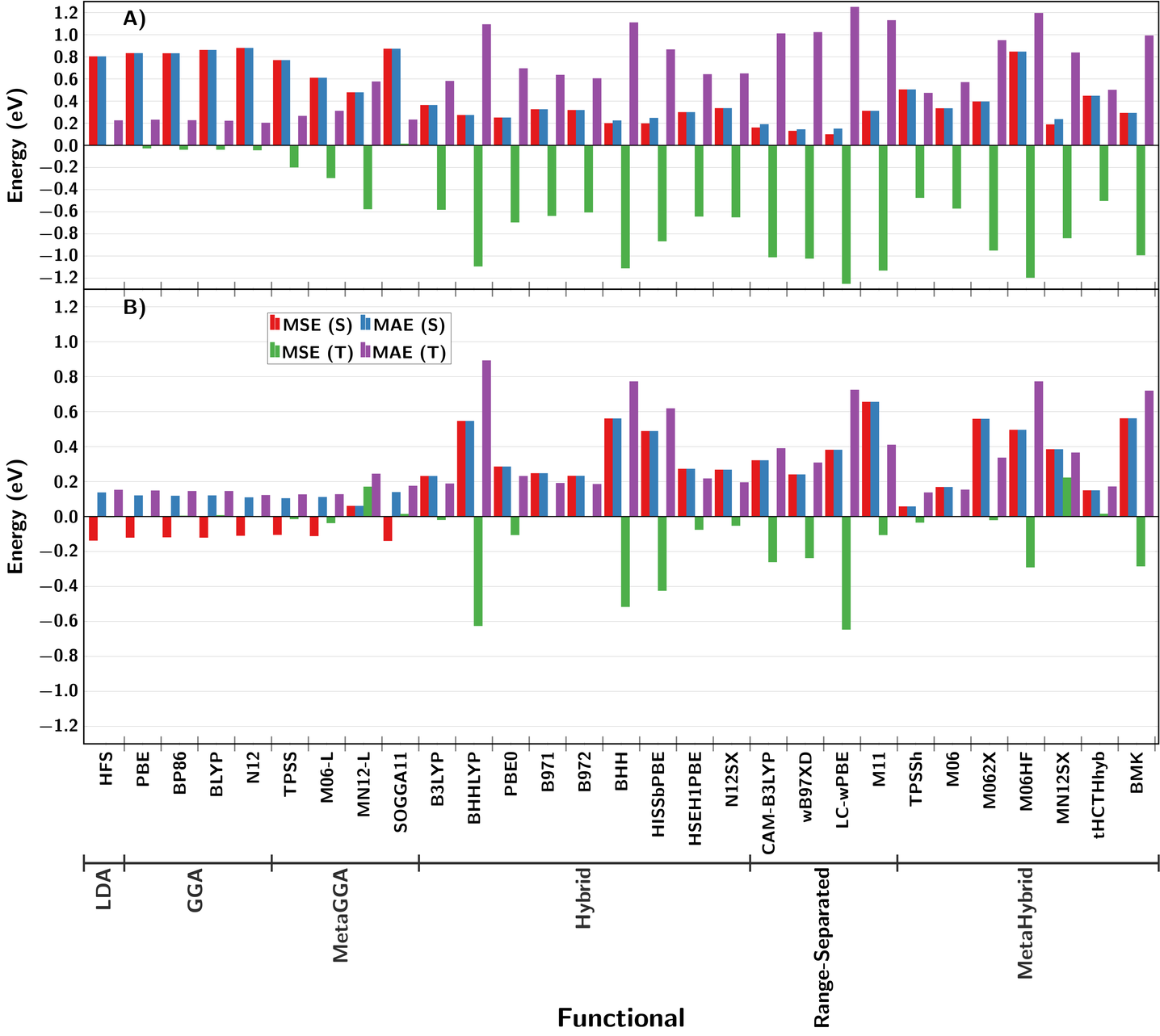}
\end{center}
\end{figure}

\begin{figure}
\caption{Mean signed errors (MSE) and mean absolute errors (MAE) of the excitation energies computed with TDA (panel A) and the energy gaps for singlets (S) and triplet (T) states (panel B) of \rupapy\ computed against MS-CASPT2(18,14). }
 \label{fig:excenrgfuncttdapapy}
\begin{center}
 \includegraphics[width=\textwidth]{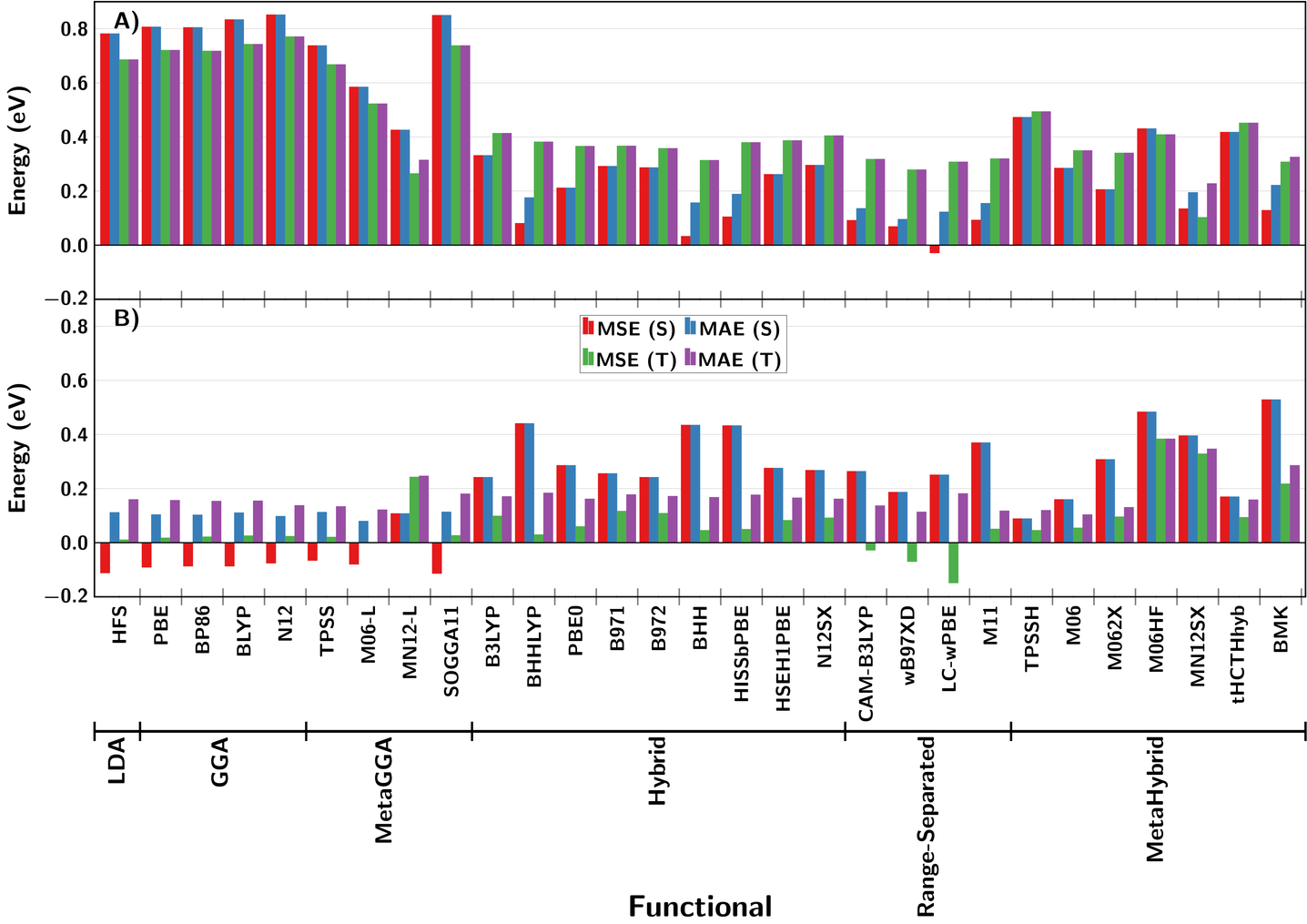}
\end{center}
\end{figure}

Figures \ref{fig:excenrgfunctpapy}A and \ref{fig:excenrgfuncttdapapy}A show the MSE and MAE of the vertical excitation energies calculated with all 29 XC functionals with respect to the MS-CASPT2(18,14) results, for the singlet and triplet states without and with TDA, respectively.
One can observe that LDA and GGA's provide almost identical results, strongly underestimating the singlet energies. This underestimation is at least systematic as MSE and MAE are of the same size.
For the triplet excitation energies obtained with LDA and GGA's the MSE and MAE are much smaller than for the singlets; however, this error is not systematic.

In the metaGGA XC functionals, with the exception of SOGGA11, the MSE and MAE for the singlets is reduced whilst for the triplets is increased, albeit the errors for the triplets become more systematic than in the case of the non-meta GGA XC functionals without TDA.

The hybrid XC functionals and their meta and range-separated counterparts typically show  smaller MSE and MAE for the singlets than the (meta)GGA XC functionals, with the exception of M06HF.
For the triplets, the observation is the opposite, with all of the hybrid XC functionals providing larger MSE's and MAE's compared to the GGA XC functionals. However, compared to the GGA XC functionals the errors for the triplets are systematic.

With the use of TDA the observations for the singlets remain mostly unaltered.
For the LDA and GGA XC functionals the errors are still systematic and of the same order of magnitude as without TDA. Also for the metaGGA functionals the same reduction in the errors for the singlets is seen. However, the MSE and MAE of the triplets are slightly smaller than that of the singlets and more importantly have become systematic.
The hybrid XC functionals and their range-separated and meta counterparts show reduced MSE and MAE for the singlets, which is most noticeable for M06HF and M11.
For the triplets the error reduction is even larger across the board of hybrid XC functionals tested.
One important point is that the systematic nature of the errors for some of the hybrid functional is lost, i.e., BHHLYP, BHH, LC-wPBE, MN12SX, BMK, especially in the singlets.

The initial conclusion to draw from this data is similar to that for \rubpy, i.e. that the hybrid XC functionals using TDA show the best performance compared to the MS-CASPT2 to calculate vertical excitation energies.
Particularly suitable for \rupapy\ seem to be wB97XD for the singlet excited states and MN12SX for the triplet excited states.

The errors in the energy gaps relative to the S$_{1}$ are shown in Figure \ref{fig:excenrgfunctpapy}B and \ref{fig:excenrgfuncttdapapy}B.
As expected from the other two compounds investigated, the trends for the XC functionals are different than those found for the vertical excitation energies.
The GGA XC functionals once again show a reduced error in the energy gaps with all having an MSE and MAE of less than 0.2 eV for both the singlets and triplets, with the exception of the metaGGA XC functional MN12-L, which has an MAE larger than 0.2 eV for the triplets.
The hybrid XC functionals on average show a larger error for both the singlet and triplet excited state energy gaps than the GGA and LDA XC functionals. For the singlet excited state energy gaps all hybrid XC functionals give an MSE and MAE of over 0.2 eV, with the exception of TPSSH, M06 and tHCTHhyb.
For the triplet states energy gaps, 11 of the 20 tested hybrid XC functionals provide an MSE and MAE 2 to 3 times larger than the GGA and LDA XC functionals, whilst the other 9 provide MAE values \textit{ca.}~0.05 eV larger than the reported MAE values for the GGA and LDA XC functionals. However, for the triplet excited state energy gaps a main observation is that rarely the MSE and MAE are the same size, which indicates that the errors in the energy gaps when not using TDA are not systematic.

Upon inclusion of TDA, the errors are reduced for the GGA XC functionals and in particular for the triplet  states calculated when using a hybrid functional.
Furthermore, the inclusion of TDA has overall made the errors for the energy gaps more systematic.
Nevertheless for the GGA and LDA XC functionals the MSE is much smaller than the MAE for the triplet states.
Moreover, there are hybrid XC functionals which perform equally and in some cases slightly better for the energy gaps than the GGA XC functionals, namely TPSSH, which has a MSE and MAE smaller for both the singlet and triplet excited state energy gaps.
In 16 out of the 20 tested hybrid XC functionals the error in the singlet excited state energy gaps is 2 to 3 times larger than the MSE and MAE for the GGA XC functionals.

From these results, same as for \rubpy, one can conclude that the use of standard GGA type XC functional, such as PBE or BP86, should  be more suitable to describe excited state dynamics than the hybrid XC functionals.

\subsubsection{State Character}
Finally, here we analyze the state character using BP86/TDA as a baseline.
As in previous complexes, Table  \ref{tab:StaCharSpapy} lists the main orbital transitions that contribute to the state for both MS-CASPT2 and BP86/TDA.
The corresponding BP86/TDA orbitals are displayed in Figure S5.
As discussed for the \rubpy\ and \rupy\ BP86/TDA
shows an overall agreement for the main characters of the orbitals involved in the transitions compard to MS-CASPT2, with only some further delocalisation of the orbitals over the molecule.

When comparing the calculated transitions directly the initial impression is that the BP86/TDA does not reproduce the MS-CASPT2 results for the singlet excited states. In the S$_{1}$ BP86/TDA predicts a single transition originating from composite orbitals and ending in an antibonding orbital of the metal-NO ligand. This could be generalised as a mixture of MLCT and Ligand-to-Ligand-Charge-Transfer (LLCT) transitions. MS-CASPT2 has a contribution to the S$_1$ which corresponds to the calculated transition with BP86/TDA, however the major contributor comes from the $d_{xy}$ and goes to the same antibonding orbital as calculated with BP86/TDA. In general if one considers the excitation type than in MS-CASPT2 the transitions are all of MLCT character with some LLCT charatcer mixed in.  In this regard, BP86/TDA can be considered to qualitatively reproduce the MS-CASPT2 excitations for the singlet states \cite{freitag_theoretical_2014}. Nevertheless on a one-to-one comparison only the S$_{3}$ and S$_{4}$ could be considered to be partially reproduced by the BP86/TDA calculation.

For the triplet excited states the correspondence  between MS-CASPT2 and BP86/TDA improves, with the main excitation character of the T$_{1}$ and T$_{2}$ being similar. However, for the higher triplet excited states one can only conclude the same as for the singlet excited states, that the BP86/TDA qualitatively reproduces the series of excitations being a combination of MLCT and LLCT transitions but does not give the same one-to-one like correspondence observed for the orbital transitions calculated for \rupy\ and \rubpy.

\begin{table}
\caption{Comparison of primary character of excitations within MS-CASPT2$^a$ and BP86 with TDA for the lowest 4 singlet and 6 triplet excitations of \rupapy}
\label{tab:StaCharSpapy}
\begin{center}
{\footnotesize
\begin{tabular}{ccccc}
\hline\hline
       \multicolumn{5}{c}{State Character}\\
 State& MS-CASPT2& Weight & BP86/TDA & Weight \\
\hline
\multirow{2}{*}{S$_1$}&$d_{xy}\rightarrow\pi_x^*-d_{xz}$&0.58&$d_{yz}+\pi_{y}^*+n_{N amide}+n_{O amide}\rightarrow\pi_{x}^*-d_{xz}$&0.98\\
&$d_{yz}+\pi_{y}^*+n_{N amide}\rightarrow\pi_x^*-d_{xz}$&0.22&&\\[4pt]
\multirow{2}{*}{S$_2$}&$d_{xy}\rightarrow\pi_y^*-d_{yz}$&0.76&$d_{yz}+\pi_{y}^*+n_{N amide}+n_{O amide}\rightarrow\pi_{x}^*-d_{xz}$&0.56\\
&&& $d_{yz}+\pi_{y}^*+n_{N amide}+n_{O amide}\rightarrow\pi_{y}^*-d_{yz}$&0.37\\[4pt]
\multirow{2}{*}{S$_3$}&$d_{yz}+\pi_{y}^*+n_{N amide}\rightarrow\pi_x^*-d_{xz}$&0.59& $d_{yz}+\pi_{y}^*+n_{N amide}+n_{O amide}\rightarrow\pi_{y}^*-d_{yz}$&0.66\\
&$d_{xy}\rightarrow\pi_x^*-d_{xz}$&0.21& $d_{yz}+\pi_{y}^*+n_{N amide}+n_{O amide}\rightarrow\pi_{x}^*-d_{xz}$&0.23\\[4pt]
\multirow{3}{*}{S$_4$}&$d_{yz}+\pi_{y}^*+n_{N amide}\rightarrow\pi_y^*-d_{yz}$&0.48&$d_{yz}+\pi_{y}^*+n_{N amide}+n_{O amide}\rightarrow\pi_{y}^*-d_{yz}$&0.24\\
&$d_{xz}+\pi_x^*\rightarrow\pi_x^*-d_{xz}$&0.31&$d_{xy}\rightarrow\pi_{y}^*-d_{yz}$&0.23\\
&&&$d_{yz}+\pi_{y}^*+n_{N amide}+n_{O amide}\rightarrow\pi_{y}^*-d_{yz}$&0.17\\
\cmidrule{1-5}
\multirow{2}{*}{T$_1$}&$d_{yz}+\pi_{y}^*+n_{N amide}\rightarrow\pi_x^*-d_{xz}$&0.35&$d_{yz}+\pi_{y}^*+n_{N amide}+n_{O amide}\rightarrow\pi_{y}^*-d_{yz}$&0.66\\
&$d_{xy}\rightarrow\pi_x^*-d_{xz}$&0.30& $d_{yz}+\pi_{y}^*+n_{N amide}+n_{O amide}\rightarrow\pi_{x}^*-d_{xz}$&0.17\\[4pt]
\multirow{2}{*}{T$_2$}&$d_{yz}+\pi_{y}^*+n_{N amide}\rightarrow\pi_y^*-d_{yz}$&0.53&$d_{yz}+\pi_{y}^*+n_{N amide}+n_{O amide}\rightarrow\pi_{x}^*-d_{xz}$&0.74\\
&&&$d_{yz}+\pi_{y}^*+n_{N amide}+n_{O amide}\rightarrow\pi_{y}^*-d_{yz}$&0.22\\[4pt]
T$_3$&$d_{xy}\rightarrow d_{x^2-y^2}$&0.53&$d_{yz}+\pi_{y}^*+n_{N amide}+n_{O amide}\rightarrow\pi_{x}^*-d_{xz}$&0.79\\[4pt]
\multirow{2}{*}{T$_4$}&$d_{xy}\rightarrow\pi_x^*-d_{xz}$&0.36&$d_{yz}+\pi_{y}^*+n_{N amide}+n_{O amide}\rightarrow\pi_{y}^*-d_{yz}$&0.81\\
&$d_{yz}+\pi_{y}^*+n_{N amide}\rightarrow\pi_x^*-d_{xz}$&0.21&&\\[4pt]
T$_5$&$d_{xy}\rightarrow\pi_y^*-d_{yz}$&0.51&$d_{xy}\rightarrow\pi_{x}^*-d_{xz}$&0.99\\[4pt]
T$_6$&$d_{xz}+\pi_x^*\rightarrow\pi_x^*-d_{xz}$&0.83&$d_{xy}\rightarrow\pi_{y}^*-d_{yz}$&0.98\\
\hline\hline
\end{tabular}\\
$^a$ Characters for MS-CASPT2 calculation from Ref.~\citenum{freitag_theoretical_2014}.}
\end{center}
\end{table}

Though we will use the BP86/TDA as a baseline for comparing other XC functionals due to its qualitative agreement, one cannot categorically conclude that any state reordering occurring with the other functionals is necessarily incorrect, especially for the singlet excited states. For the triplet excited states one can use the T$_{1}$ and T$_{2}$ as a judge of how well the other functionals perform as BP86/TDA reproduces satisfactorily the orbital transitions observed in MS-CASPT2.

\begin{figure}
\caption{Wave function overlap of the singlet (A) and triplet (B) states of \rupapy\ calculated using different functionals with respect to BP86.}
 \label{fig:papy_overl}
\begin{center}
 \includegraphics[width=\textwidth]{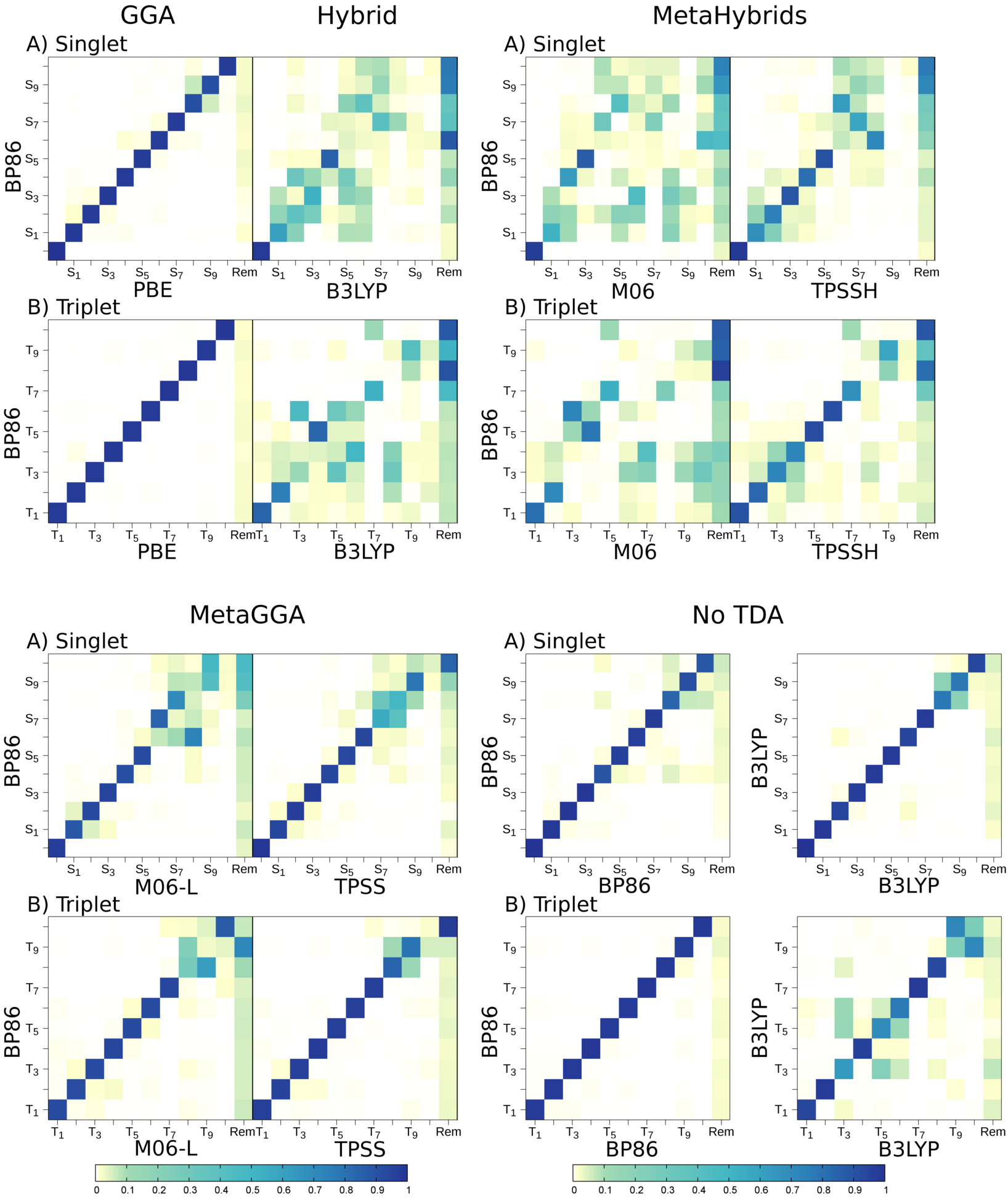}
\end{center}
\end{figure}

Figure \ref{fig:papy_overl} shows the wavefunction overlaps for selected XC functionals, as well as a comparison of the effect of TDA on the wavefunction for BP86 and B3LYP.
As previously observed for both \rupy\ and \rubpy\, using PBE instead of BP86 has negligible effects on the singlet excited states with only a slight mixing of the S$_{8}$ and S$_{9}$ and no effect on the triplet states.

In the case of B3LYP one can see that only the ground state has a large overlap with that provided by BP86/TDA. All the calculated singlet excited states have significant mixing, whilst those from the S$_{6}$ and higher have larger amounts of the BP86/TDA state character missing. For the triplet states, which in the case of \rupapy\ provide a better baseline for judging the functional, one observers that B3LYP reproduces the T$_{1}$ and most of the T$_{2}$ state character. The higher excited state triplets show significant mixing with the T$_{7}$ onwards mostly not being present in the excitation window of BP86/TDA.

M06 and TPSSH have been chosen as a representative of metaHybrid XC functionals.
M06 shows almost identical behaviour as to that of B3LYP for both singlet and triplet excited states, whilst TPSSH  shows less mixing of the states up to the S$_{5}$ and T$_{7}$, thus reproducing more exactly the excited states.

The calculated overlaps for two metaGGA XC functionals (M06-L and TPSS) provide for the lower excited states (up to S$_{5}$ and T$_{7}$) the same behaviour as PBE.
From the S$_{6}$ and T$_{8}$ upwards, however, M06-L shows more state mixing and also larger amounts of state character from the BP86/TDA calculation not being reproduced.
For TPSS this only occurs from the S$_{7}$ upwards but the S$_{10}$ and T$_{10}$ from BP86/TDA are for the most part not reproduced by TPSS.

Finally, it is interesting to consider the effect of TDA on the state characters calculated for \rupapy. With BP86 the effect of TDA is small, with only the S$_{7}$ upwards showing any mixing when not using TDA.
For the triplet excited states, these remain identical irrespective of TDA.
In B3LYP, the use of TDA has a smaller effect on the singlet excited states than that observed with BP86. However, for the triplet states more mixing of the excited states is observed in the absence of TDA.
Nevertheless, the state ordering remains the same up to the T$_{8}$ with
only the T$_{9}$ and T$_{10}$ switching places in terms of state order. From this comparison one could conclude that the use of TDA in the case of \rupapy\ is not necessary as the state ordering is not severely affected. However, it is still recommended from the effect it had on the energetics of the excited states, rather than its effects on the excited state characters.

\section{Conclusions}

While in many literature benchmark studies, the performance of a TD-DFT is assessed with regards to its accuracy in the calculation of the vertical excitation energies with respect to the ground state energy, in many applications it is also required that energy gaps with respect to a state different from the ground state are also adequately described.
For instance, the prediction of molecular mechanisms using nonadiabatic molecular dynamics methods is very sensitive to energy gaps between excited states, as the coupling between two (or more) states is determined by their energy gap and so nonadiabatic population transfer depends on the energy difference between the involved states.
Certainly, the shape and slope of the potential energy surfaces can also influence the outcome of photodynamics simulations and in doubtful cases, it can be useful to explore the performance of different functionals to describe relaxation pathways, as e.g. done in Ref.\citenum{valsson_regarding_2015}.
With dynamics applications in mind and since in general, the gradients of the potential energy surfaces are difficult to analyze systematically, in this paper we have focused on the analysis of the energy gaps between the electronic excited states of three Ru(II) complexes and their associated wavefunction characters using different density functionals.
As a reference, the results of a multiconfigurational MS-CASPT2 calculation was employed.
We found that the trends observed in the accuracy of the excitation energies are different from those observed for the energy gaps between excited states and therefore, the common assessment of a functional should also include a benchmark of the energy gaps.
Hence, when considering excitation energies from the electronic ground state, we found that the hybrid functionals and their range-separated and meta counterparts are the most accurate, while the best energy gaps between the electronic excited states tend to be best described by the pure GGA exchange-correlation functionals.
In \rubpy\ and \rupapy\ the range of GGA functionals tested provided similar errors for the excited state energy gaps, typically half of what was given by the majority of the hybrid functionals.
In contrast, in \rupy\ there is not a clear distinction on the performance of the functionals.
The excited state energy gaps of the singlet excited states are all described similarly with all the functionals.
However, for the triplet states the GGA functionals perform predominantly better than the family of hybrid functionals.
A noticeable exception is M06HF, which gives errors for the triplet state energy gaps as small as with the pure functionals.

Particular attention was put to investigate the effect of the functional on the character of the  electronic wavefunction in each state.
While it is possible to compare the composition of each state by examining one-by-one the determinant contributions and orbitals involved, here a more efficient procedure based on wavefunction overlaps was employed to systematically evaluate changes in the state character of the states.
Since BP86 showed the best state ordering and state mixing in comparison to the multiconfigurational MS-CASPT2 reference, it was taken as a baseline to compare all other functionals.
In general, we found that all pure GGA and metaGGA functionals performed similarly,
while the hybrid functionals showed a significant mixing with respect to the BP86 functional.
In particular, the deviations of the hybrid B3LYP are more noticeable for the higher energy singlet states with some states not described at all.
The triplet states also show dramatic state reordering with B3LYP.
MetaHybrids deliver the same behaviour as B3LYP for the singlet and triplet states.
Assuming that the order and character predicted by MS-CASPT2 is correct, one is left to conclude that hybrid and metaHybrid functionals are not adequate to describe the electronic excited states of Ru polypyridyl complexes; instead, pure functionals, such as BP86 or PBE or metaGGAs, such as M06-L or TPSS, might be more adequate, in particular, if one is interested to calculate non-radiative properties or perform non-adiabatic dynamics.
For properties that depend on vertical excitation energies, such as absorption spectra or luminescence spectra, hybrid functionals are better options for the Ru(II) complexes investigated and this should in principle hold for other Ru polypyridyl complexes.

Besides the choice of the density functional, the effect of the basis set, solvation model, and different technical details on the energetics were analyzed.
From all of them, the more significant changes were found in some of the complexes with TDA, which is recommended to use.

\begin{acknowledgement}
We would like to thank the Austrian Science Fund (FWF) within the project M 1815-N28 and the ITN-EJD (TCCM) for financial support. We thank Maximilian Menger for advising how to acquire the information required for calculating wavefunction overlaps within Gaussian09. Part of the calculations have been carried out in the Vienna Scientific Cluster (VSC).
\end{acknowledgement}

\begin{suppinfo}
A comparison of the excitation energies calculated with the geometry optimized at different levels of theory is given in the SI for \rubpy. The second section gives a more detailed discussion on the effect of the technical details of the calculation on the excited state energies. The third
section considers the effect of solvation and technical details on the excitation energies calculated using hybrid functionals. Section IV shows the primary orbitals involved in the excitations calculated at the BP86 level. Section V shows wavefunction overlap pictures of other functionals
for \rubpy\ and for an example of solvation with B3LYP. The sixth and final section discusses the effect of the technical details for B3LYP and shows the wavefunction overlaps of the technical details for both BP86	 and B3LYP.  This material is available free of charge via the Internet at
\texttt{http://pubs.acs.org/}.
\end{suppinfo}

\bibliography{Zotero}

\end{document}